\documentclass[11pt,a4paper]{article}
\usepackage[T1]{fontenc}
\usepackage{jheppub}
\usepackage{amsmath}
\usepackage{amssymb}
\usepackage{mathtools}
\usepackage{latexsym}
\usepackage{bm}
\usepackage{eufrak}
\usepackage{bbold}         
\usepackage{braket}
\usepackage{slashed}
\numberwithin{equation}{section}

\usepackage{graphicx}
\usepackage{booktabs}
\usepackage{multirow}
\usepackage{framed}

\usepackage{tikz}
\usetikzlibrary{
    calc,
    arrows.meta,
    patterns,
    decorations.pathreplacing,
    shapes.misc
}

\usepackage{xcolor}
\definecolor{darkgreen}{RGB}{0,100,0}
\definecolor{darkpink}{RGB}{199,30,110}

\usepackage{stackengine}    %% needed by \reallywidetilde below
\usepackage{scalerel}       %% needed by \reallywidetilde below
\usepackage[normalem]{ulem} %% normalem: keeps \emph italic, not underlined

\usepackage{hyperref}
\usepackage{upgreek}

\newcommand{\Tr}{\operatorname{Tr}}

\makeatletter
\newcommand{\doublewidetilde}[1]{{%
  \mathpalette\double@widetilde{#1}%
}}
\newcommand{\double@widetilde}[2]{%
  \sbox\z@{$\m@th#1\widetilde{#2}$}%
  \ht\z@=.9\ht\z@
  \widetilde{\box\z@}%
}
\makeatother

\author{
Audrey Lindsay${}^{1}$ ,\, Tomasz R.\ Taylor${}^{1,2}$\\[0.5cm]
 $^1${\it Department of Physics,
  Northeastern University, Boston, MA 02115, USA}\\
  $^2${\it Faculty of Physics, University of Warsaw, Pasteura 5, 02-093 Warsaw, Poland}\\[0.2cm]
}

\emailAdd{a.lindsay@northeastern.edu}
\emailAdd{taylor@neu.edu}
\title{\boldmath {Gluon Scattering in Expanding de Sitter Universe} \unboldmath}

\abstract{We construct an observer-centric S-matrix in the expanding Poincar\'e patch of four-dimensional de Sitter space. The asymptotic states are defined along the worldline of a comoving, or \emph{eternal}, observer, the
only kind of geodesic observer that remains in the patch for all proper time, and the free Hamiltonian is the conserved charge generating translations in her proper time. The resulting S-matrix is independent of which eternal observer is chosen. We apply the construction to pure Yang--Mills theory in Coulomb gauge and compute the three- and four-gluon
amplitudes at tree level. The amplitudes have a structure similar to the cosmological correlators of gauge currents and are constrained by the Ward identities of de Sitter isometries.}

\gdef\@fpheader{}
\makeatother

\begin{document}
\maketitle
\section{Introduction}

Scattering amplitudes are the natural observables of quantum field theory in flat spacetime.
They are defined by the behavior of states at asymptotically early and late times, and the
symmetries of the spacetime constrain them via Ward identities. Central to the construction is a
global time-translation symmetry: its conserved charge defines the energy, generates the evolution
between the asymptotic states, and fixes the time in which those states are specified. De Sitter
space has no such symmetry~\cite{Spradlin:2001pw, Bousso:2004tv}. There is
no globally timelike Killing vector, so there is no conserved energy for the spacetime as a whole,
the Hamiltonian that evolves states in a given time coordinate is in general time dependent, and
there is no preferred time in which to specify in- and out-states. The problem is compounded by
the cosmological horizon that surrounds every observer and by the spacelike character of the
asymptotic boundaries $\mathcal{I}^\pm$. The flat-space construction therefore does not carry over directly.

While there is no canonical S-matrix in de Sitter space, different slicings make different structures manifest, and each has been used to understand a different aspect of quantum field theory there. The static patch is the region causally accessible to a single observer, bounded by
her cosmological horizon. Restricted to it, the Bunch--Davies state~\cite{Bunch:1978yq} is thermal at the
Gibbons--Hawking temperature $H/2\pi$, and the horizon carries an entropy~\cite{Gibbons:1977mu}. More recently, the static patch has been the setting for algebras of observables dressed to an observer's worldline~\cite{Chandrasekaran:2022cip}, and for scattering between an observer's past and future horizons, including tree-level Yang--Mills~\cite{Albrychiewicz:2020ruh, Albrychiewicz:2021ndv}. The flat slicing covers the expanding Poincar\'e patch,
half of global de Sitter space, with spatially flat sections, and it is here that de Sitter
space meets cosmology. During inflation the universe is approximately de Sitter in this slicing,
and the primordial fluctuations generated then, later imprinted on the cosmic microwave
background and the large-scale structure, are described by correlation functions on the future
boundary of the patch~\cite{Baumann:2009ds, Chen:2010xka}. The observables in this setting are these late-time correlators, or equivalently the
wavefunction of the universe~\cite{Maldacena:2002vr, Maldacena:2011nz}, rather than S-matrix
elements. The de Sitter isometries act on $\mathcal{I}^+$ as conformal transformations of
$\mathbb{R}^3$, so these boundary observables are constrained like correlators of a Euclidean
conformal field theory. This is the starting point of dS/CFT, which identifies the late-time
wavefunction with the partition function of such a theory
~\cite{Strominger:2001pn, Witten:2001kn} and of the cosmological bootstrap, which has
mapped out their structure in detail~\cite{Arkani-Hamed:2015bza, Baumann:2020dch, Arkani-Hamed:2018kmz, Arundine:2026fbr, Sleight:2019hfp}, including the constraints that gauge invariance places on the couplings of massless spinning fields~\cite{Sleight:2021iix}. 

S-matrices have also been constructed in the Poincar\'e patch by an LSZ reduction of
time-ordered correlators, and the late-time wavefunction coefficients and in-in correlators can
be written explicitly in terms of their elements~\cite{Melville:2023kgd, Melville:2024ove}.
In a related approach, it was shown that in-in correlators of non-dissipative systems can be
computed in the in-out formalism, using only time-ordered Feynman
propagators~\cite{Donath:2024utn}. In de Sitter space this requires extending the expanding
Poincar\'e patch by a contracting patch, and the resulting in-out formulation leads naturally to
an S-matrix. Finally, the closed slicing covers all of de Sitter space, with $S^3$ spatial sections that contract and re-expand between $\mathcal{I}^-$ and $\mathcal{I}^+$, making it the natural setting for an S-matrix connecting the two~\cite{Marolf:2012kh, Spradlin:2001nb}. Such
an S-matrix was constructed in~\cite{Taylor:2024vdc} by linking the Hilbert space to an inertial
geodesic observer, with amplitudes given by a generalized Dyson formula, and has since been
applied to scalars~\cite{Taylor:2025spp} and to the three-gluon amplitude~\cite{Taylor:2026ase};
the constraints of the $SO(1,4)$ isometries on these amplitudes were derived
in~\cite{Lindsay:2025chz}.

In this work we follow the technique of~\cite{Taylor:2024vdc} to construct an observer-centric S-matrix in the expanding Poincar\'e patch of four-dimensional de Sitter spacetime, and apply it to pure Yang--Mills theory. In Section~\ref{sec:geometry} we describe the
expanding Poincar\'e patch and its isometries, and classify its timelike geodesics, identifying
the eternal observers and the Killing vectors that generate their proper time. In
Section~\ref{sec:freefields} we turn to Yang--Mills theory. Since it is classically conformally
invariant and the patch is conformally flat, the free mode functions and interaction vertices are
those of Minkowski space, and the de Sitter structure enters only through the action of the
isometries on the modes and through the range of the conformal-time integral. We quantize in Coulomb gauge and construct the conserved charges of all ten
isometries, including the Hamiltonian of an eternal observer. The boosts do not preserve Coulomb
gauge, and act on the fields together with a compensating gauge transformation. In Section~\ref{sec:Smatrix} we construct the S-matrix from this
Hamiltonian, show that it is independent of the choice of eternal observer, and compare it with
the Bunch--Davies S-matrix of~\cite{Melville:2023kgd}. We then compute the three- and four-gluon
S-matrix elements at tree level, express them in spinor-helicity variables adapted to de Sitter
space~\cite{David:2019mos}, and compare them with the corresponding wavefunction coefficients. 
Up to the normalization of the external states, the contact contributions agree. The exchange contributions differ because the exchanged gluon is described by different propagators: the Feynman propagator for the S-matrix, and a propagator that vanishes at $\mathcal{I}^+$ for the wavefunction. Section~\ref{sec:WI} derives the Ward identities, the Poincar\'e-patch
counterparts of those obtained in global de Sitter space
in~\cite{Lindsay:2025chz}. Translations, rotations, and the dilation preserve Coulomb gauge, so their Ward identities are homogeneous. The boosts do not preserve Coulomb gauge, and the compensating gauge transformation required to restore the field to the gauge slice is the source of an
inhomogeneous term in their Ward identity. This term is not an artifact of quantizing in Coulomb gauge: the same term appears in the special conformal
Ward identity for boundary correlators of non-abelian
currents~\cite{Maldacena:2011nz}, where it follows from current conservation
and no gauge is fixed. We derive this identity explicitly at three points, where
the boost becomes, up to the normalization of the external states, the
momentum-space special conformal generator~\cite{Bzowski:2013sza,
Witten:2003nn}, and show that the corresponding term is nonzero at four
points. In Section~\ref{sec:flatlimit} we turn to the
flat limit. For a conformal theory in a conformally flat patch the de Sitter radius drops out, so
there is no parameter in which to take the limit. Instead, the time integral ending at $\mathcal{I}^+$ produces $i/(k_T+i\epsilon)$ where
Minkowski space would give $2\pi\delta(k_T)$, and the flat-space amplitudes, from the three-gluon
amplitudes to the four-gluon MHV amplitude~\cite{Parke:1986gb}, appear as residues of this
total-energy pole~\cite{Maldacena:2011nz, Raju:2012zr, Farrow:2018yni}. We discuss in which channels the pole can be approached with real momenta, and show that the boost Ward identity is consistent with the flat limit. We conclude in Section~\ref{sec:conclusions}, and collect our spinor-helicity conventions in
Appendix~\ref{app:spinorhelicity}.

\section{De Sitter geometry, geodesics, and observers}\label{sec:geometry}

\begin{figure}
\centering
\begin{tikzpicture}[scale=6]

\fill[pink, opacity=0.3] (1,0) -- (0,1) -- (1,1) -- cycle;
\draw[thick, darkpink] (0.997, 0) -- (0.997,1);

\node[above] at (0.5,1) {$\mathcal{I}^+$};
\node[below] at (0.5,0) {$\mathcal{I}^-$};
\node[left]  at (0,0.5) {S};
\node[right] at (1,0.5) {N};
\draw[very thick] (0,0) rectangle (1,1);
\draw[very thick] (0,1) -- (1,0);
\end{tikzpicture}
\caption{Penrose diagram of de Sitter spacetime. Each point represents an $S^2$,
except on the vertical edges, which are the worldlines of the south (S) and north (N)
poles of the spatial $S^3$. The shaded
upper-right triangle is the expanding Poincar\'e patch, corresponding to
$X^0 + X^4 > 0$; it contains the entire worldline of the north pole.
The unshaded lower-left triangle is the contracting Poincar\'e patch, covering
$X^0 + X^4 < 0$. It is described by the same metric~\eqref{eq:ppmetric}, but with
$\eta \in (0, \infty)$. $\mathcal{I}^{\pm}$ are the future and past spacelike boundaries.}
\label{fig:penrose}
\end{figure}
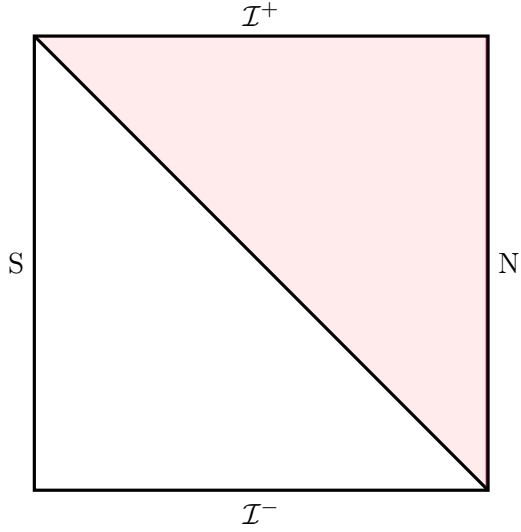

Four-dimensional (global) de Sitter spacetime ${\rm dS}_4$ is realized as the Lorentzian hyperboloid of radius $\ell = H^{-1}$, with $H$ the Hubble rate, embedded in Minkowski space $\mathbb{R}^{1,4}$ with metric $\eta_{AB} = {\rm diag}(-,+,+,+,+)$:
\begin{equation}\label{eq:hyperboloid}
-X_0^2 + X_1^2 + X_2^2 + X_3^2 + X_4^2 = \ell^2.
\end{equation}
It is a solution of the vacuum Einstein equations with cosmological constant $\Lambda = 3/\ell^2$, and the isometry group is the subgroup of $O(1,4)$ preserving~\eqref{eq:hyperboloid}, generated by the ten Killing vectors
\begin{equation}\label{eq:MAB}
M_{AB} = X_A\,\partial_B - X_B\,\partial_A .
\end{equation}
Global ${\rm dS}_4$ is covered by the closed  slicing,
\begin{equation}\label{eq:globalmetric}
    ds^2 = -dT^2 + \ell^2\cosh^2\!\left(T/\ell\right)d\Omega_3^2,
    \qquad
    X^0 = \ell\sinh\!\left(T/\ell\right),\quad X^{a} = \ell\cosh\!\left(T/\ell\right)\Omega^{a},
\end{equation}
with $a = 1,\dots,4$ and $\Omega^a\Omega_a = 1$, which presents ${\rm dS}_4$ as a spatial $S^3$ that contracts to radius $\ell$ and re-expands. Setting $\sec\mathcal{T} = \cosh(T/\ell)$ compactifies the time coordinate to $\mathcal{T}\in(-\tfrac\pi2,\tfrac\pi2)$ and brings~\eqref{eq:globalmetric} to
\begin{equation}
ds^2 = \ell^2\sec^2\!\mathcal{T}\,(-d\mathcal{T}^2+d\Omega_3^2),
\end{equation}
representing ${\rm dS}_4$ as conformal to a finite strip of the Einstein static universe $\mathbb{R}\times S^3$, from which the Penrose diagram in Fig.~\ref{fig:penrose} is read off.
 
Different charts on ${\rm dS}_4$ cover different portions of the diagram and make different
structures manifest. We will work in the expanding Poincar\'e patch, which covers half of global
${\rm dS}_4$ and foliates that half by spatially flat $\mathbb{R}^3$ slices; as discussed in the
introduction, it is the chart relevant for cosmology.

\subsection{The expanding Poincar\'e patch}

The metric of the expanding Poincar\'e patch is
\begin{equation}\label{eq:ppmetric}
    ds^2 = \frac{\ell^2}{\eta^2}\left(-d\eta^2+d{\bf x}^2\right),
\end{equation}
with conformal time $\eta\in(-\infty,0)$ and $x^i\in(-\infty,\infty)$, related to the embedding coordinates by
\begin{equation}\label{eq:embeddingpoincare}
X^0  = \frac{\ell}{2\eta}\left(\eta^2-{\bf x}^2 -1\right), \quad X^i  = -\frac{\ell x^i}{\eta}\,,\quad 
X^4  = -\frac{\ell}{2\eta}\left(\eta^2-{\bf x}^2 +1\right).
\end{equation}
For simplicity, we will set $\ell = 1$ from here. The symmetry generators $M_{AB}$ in these coordinates are
\begin{equation}\label{eq:ppMABs}
\begin{aligned}
P_i & \equiv M_{i0} - M_{i4} = \partial_i\,,\\
J_i & \equiv \tfrac12\epsilon_{ijk}M_{jk} = \epsilon_{ijk}x_j\partial_k\,,\\
D & \equiv -M_{04} = -\eta\,\partial_{\eta} - x_i\,\partial_i\,,\\
K_i & \equiv M_{i0}+M_{i4} = -2\,\eta\, x_i \,\partial_{\eta} + (-\eta^2 + {\bf x}^2)\,\partial_i - 2\,x_i\,x_j\,\partial_j\,,
\end{aligned}
\end{equation}
which we have reorganized in a way natural for the flat slicing as translations $P_i$, rotations \nolinebreak$J_i$, a dilation $D$, and boosts $K_i$. Their commutation relations are
\begin{equation}\label{eq:PJDKcomms}
\begin{aligned}
    [J_i, J_j] &= -\epsilon_{ijk}J_k, &\qquad [J_i, P_j] &= -\epsilon_{ijk}P_k, &\qquad [J_i, K_j] &= -\epsilon_{ijk}K_k, \\[2pt]
    [D, J_i]   &= 0,                  &\qquad [D, P_i]   &= P_i,                &\qquad [D, K_i]   &= -K_i, \\[2pt]
    [P_i, P_j] &= 0,                  &\qquad [K_i, K_j] &= 0,                  &\qquad [P_i, K_j] &= 2\left(\epsilon_{ijk}J_k + \delta_{ij}D\right).
\end{aligned}
\end{equation}

The expanding Poincar\'e patch is the region $X^0+X^4 = -\eta^{-1} > 0$, and an isometry maps the patch to itself precisely when its action on $X^0+X^4$ vanishes on the horizon. Translations annihilate $X^0+X^4$ and rotations act only on $X^{1,2,3}$, while the dilation rescales it,
\begin{equation}
    P_i\left(X^0+X^4\right) = 0,\qquad J_i\left(X^0+X^4\right) = 0, \qquad D\left(X^0+X^4\right) = X^0+X^4,
\end{equation}
so each of $\{P_i, J_i, D\}$ maps the expanding patch to itself. Together they generate the similitude group
\begin{equation}\label{eq:sim3}
    {\rm Sim}(3)\cong\mathbb{R}^3\rtimes\left(SO(3)\times\mathbb{R}^+\right),
\end{equation}
 a proper subgroup of the full isometry group $SO(1,4)$. The boosts are not among them since
\begin{equation}
K_i\left(X^0+X^4\right) = 2X_i
\end{equation}
does not vanish on the horizon. The future boundary of the patch is reached as $\eta\to0^-$, where the conformal factor in~\eqref{eq:ppmetric} diverges and the slices approach $\mathcal{I}^+$. Since~\eqref{eq:ppmetric} is conformally flat, the generators~\eqref{eq:ppMABs} degenerate on $\mathcal{I}^+$ to the conformal group of $\mathbb{R}^3$.

Absent from~\eqref{eq:ppMABs} is any generator that translates conformal time: $\partial_\eta$ is not a Killing vector of~\eqref{eq:ppmetric}. Fixing a chart is therefore not the same as fixing a notion of time, and the scattering problem we pose requires the latter. In- and out-states are specified by their behavior at asymptotically early and late times along a worldline, and the Hamiltonian that evolves between them is the conserved charge of a Killing vector timelike along that worldline. Both ingredients come from an observer rather than from the chart. Since the patch is a proper subregion of ${\rm dS}_4$, a timelike curve inside it at one moment need not have been inside it in the past: a worldline that enters at finite proper time has no asymptotic past within the patch, and supports no in-state. We therefore classify the timelike geodesics of~\eqref{eq:ppmetric} by whether they remain in the patch for all proper time, and it is the ones that do that will define the S-matrix.

\subsection{Geodesic observers}

\begin{figure}[]
\centering
\begin{tikzpicture}[scale=6]

\begin{scope}
\fill[pink, opacity=0.3] (1,0) -- (0,1) -- (1,1) -- cycle;
\draw[thick, darkpink] (0.997, 0) -- (0.997,1);

\begin{scope}
\clip (1,0) -- (0,1) -- (1,1) -- cycle;

\foreach \A in {0.15,0.4,0.8,1.5,3.5} {
  \draw[dashed, thick, gray]
    plot[
      variable=\x,
      domain=0:30,
      samples=200,
      smooth
    ]
    (
      {1 - (atan(-\A+\x) - atan(-\A-\x))/180},
      {(atan(-\A+\x) + atan(-\A-\x) + 180)/180}
    );
}

\foreach \R in {0.1,0.4,0.8,1.5,3.5} {
  \draw[thick, darkpink]
    plot[
      variable=\A,
      domain=0.005:40,
      samples=210,
      smooth
    ]
    (
      {1 - (atan(-\A+\R) - atan(-\A-\R))/180},
      {(atan(-\A+\R) + atan(-\A-\R) + 180)/180}
    );
}
\end{scope}

\node[above] at (0.5,1) {$\mathcal{I}^+$};
\node[below] at (0.5,0) {$\mathcal{I}^-$};
\node[left]  at (0,0.5) {S};
\node[right] at (1,0.5) {N};
\draw[very thick] (0,0) rectangle (1,1);
\draw[very thick] (0,1) -- (1,0);
\draw[dashed, thick, gray] (0.1, 0.2) -- (0.2, 0.2);
\node[right] at (0.2, 0.2) {{\color{gray} $\eta = {\rm const.}$}};
\draw[thick, darkpink] (0.1, 0.1) -- (0.2, 0.1);
\node[right] at (0.2, 0.1) {{\color{darkpink} ${\bf x} = {\rm const.}$}};
\node[below] at (0.5,-0.12) {(a)};
\end{scope}

\begin{scope}[xshift=1.3cm]
\fill[pink, opacity=0.3] (1,0) -- (0,1) -- (1,1) -- cycle;
\fill[pink, opacity=0.6] (1,0) -- (0.5,0.5) -- (1,1) -- cycle;
\foreach \A in {0.15,0.4,0.8,1.5,3.5} {
  \draw[dashed, thick, gray]
    plot[
      variable=\x,
      domain=0:30,
      samples=200,
      smooth
    ]
    (
      {1 - (atan(-\A+\x) - atan(-\A-\x))/180},
      {(atan(-\A+\x) + atan(-\A-\x) + 180)/180}
    );
}
\node[above] at (0.5,1) {$\mathcal{I}^+$};
\node[below] at (0.5,0) {$\mathcal{I}^-$};
\node[left]  at (0,0.5) {S};
\node[right] at (1,0.5) {N};
\draw[very thick] (0,0) rectangle (1,1);
\draw[very thick] (0,1) -- (1,0);
\draw[thick] (0.5,0.5) -- (1,1);
\draw[thick, darkpink] (0.997, 0) -- (0.997,1);
\node[below] at (0.5,-0.12) {(b)};
\end{scope}

\end{tikzpicture}
\caption{Penrose diagrams of de Sitter spacetime, with the expanding Poincar\'e patch shaded.
(a) Worldlines of the eternal congruence, ${\bf x} = {\rm const.}$, and slices of
$\eta = {\rm const.}$; the observer at ${\bf x} = 0$ follows the North pole. (b) The static
patch (dark right triangle) of the observer at the North pole.}
\label{fig:penrosegeodesics}
\end{figure}
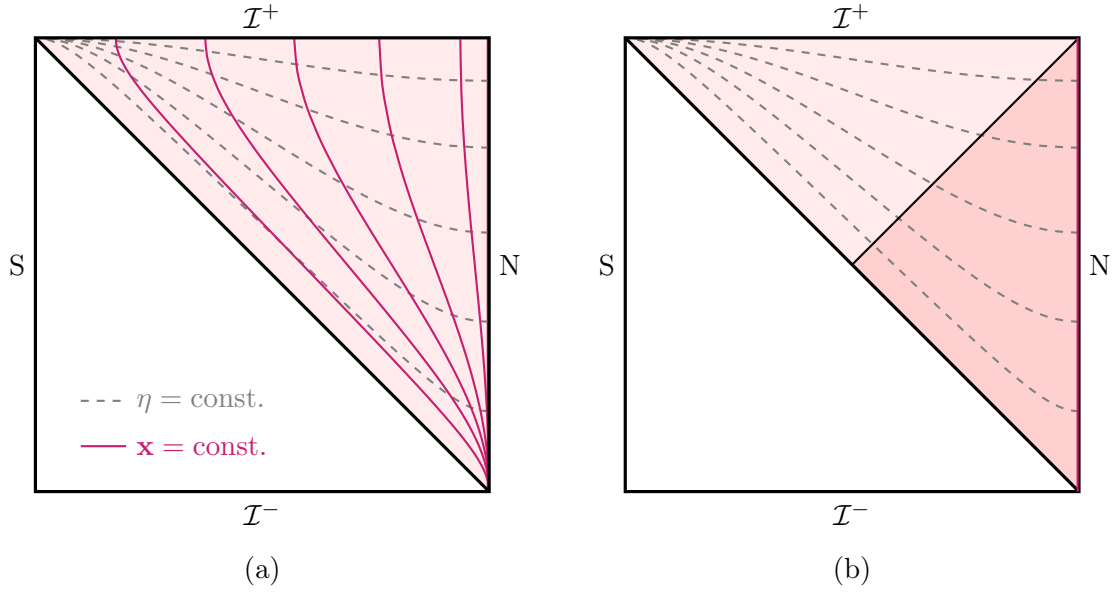

Every isometry of the patch supplies a quantity conserved along a timelike
geodesic; the charge associated with translations is what allows us to classify
them. Contracting $P_i = \partial_i$ with the tangent vector of a curve
parametrized by proper time $\tau$ gives
\begin{equation}\label{eq:pdef}
    p_i \equiv g_{\mu\nu}\,(P_i)^{\mu}\,u^{\nu} = \frac{\dot{x}_i}{\eta^2},
    \qquad u^{\mu} = \frac{dx^{\mu}}{d\tau} = \left(\dot\eta, \dot{\bf x}\right),
\end{equation}
conserved because $P_i$ is Killing. Imposing the normalization condition
\begin{equation}\label{eq:timelikegeodnorm}
    -1 = g_{\mu\nu}u^{\mu}u^{\nu} = \frac{1}{\eta^2}\left(-\dot{\eta}^2+\dot{\bf x}^2\right)
\end{equation}
gives
\begin{equation}\label{eq:etadot}
    \dot{\eta} = -\eta\sqrt{1+\eta^2{\bf p}^2},
\end{equation}
with the sign fixed by taking the geodesic to be future-directed. 

First consider the case ${\bf p}\neq0$, for
which~\eqref{eq:etadot} integrates to
\begin{equation}\label{eq:etapnot0}
\eta(\tau) = -\frac{1}{|{\bf p}|}\,{\rm csch}\left(\tau - \tau^*\right),
\qquad \tau > \tau^*
\end{equation}
with $\tau^*$ an integration constant fixed by the choice of origin for proper
time. Such a geodesic reaches $\mathcal{I}^+$ only as $\tau\to\infty$, but it
crosses the past horizon $X^0+X^4=0$ at the finite proper time $\tau^*$: its
history before $\tau^*$ lies outside the patch. Every geodesic with
${\bf p}\neq0$ therefore enters the patch from elsewhere, and has no asymptotic
past within it.

Now consider ${\bf p} = 0$. These geodesics have $\dot{\bf x} = 0$ and therefore sit at fixed comoving position ${\bf x} = {\bf a}$, with~\eqref{eq:etadot} reducing to
$\dot\eta = -\eta$ and
\begin{equation}\label{eq:etap0}
\eta(\tau) = -e^{-\tau},
\end{equation}
where the origin of proper time has been chosen to set the integration constant. As $\tau$
runs over $(-\infty,\infty)$, $\eta$ sweeps the full range $(-\infty,0)$: these
worldlines lie in the patch for their entire history, and we call them ``eternal observers.'' In the FLRW form of~\eqref{eq:ppmetric}
\begin{equation}\label{eq:FLRWmetric}
    ds^2 = -d\tau^2 + e^{2\tau}d{\bf x}^2
\end{equation}
it is clear they are the comoving
observers, with their proper time $\tau$ the same as the cosmic time. Eternal observers are labeled by their position ${\bf a}$, and their trajectories are drawn
in Fig.~\ref{fig:penrosegeodesics}. There is exactly one through each point of the patch, so
they form a congruence, and we refer to it as the eternal congruence. The same subgroup that
preserves the patch preserves this eternal congruence: rotations, translations, and the dilation all map eternal
observers to other eternal observers. The boosts do not. Acting with $K_i$ on an eternal
observer produces a geodesic with ${\bf p}\neq0$, which by~\eqref{eq:etapnot0} enters the
patch at finite proper time.

Each eternal observer has her own generator of proper time. For the observer at
${\bf x} = {\bf a}$, it is the Killing vector
\begin{equation}\label{eq:DaP}
D + {\bf a}\cdot{\bf P} = -\eta\,\partial_\eta - (x_i-a_i)\,\partial_i ,
\end{equation}
which reduces on her worldline to $-\eta\,\partial_\eta = \partial_\tau$. The corresponding
conserved charge is the energy, relative to this observer, of a particle on any timelike or
null geodesic. It coincides with the energy defined in~\cite{Cacc:2008} by contracting the
particle's conserved charges $M_{AB}$ with the two null vectors that span the plane cutting
the observer's geodesic out of the de Sitter hyperboloid. Following~\cite{Taylor:2024vdc}, we will promote these classical Noether charges to
operators, where~\eqref{eq:DaP} becomes the free
Hamiltonian generating evolution in the observer's proper time. 

\section{Free fields and conserved charges}\label{sec:freefields}

Yang--Mills in the expanding patch is a flat-space problem. Four-dimensional Yang--Mills is classically conformally invariant and the patch is conformally flat, so the free mode functions are the Minkowski ones; the de Sitter structure enters only through the action of the isometries~\eqref{eq:ppMABs} on those modes.

The action for pure Yang--Mills on~\eqref{eq:ppmetric} is
\begin{equation}\label{eq:YMaction}
S = -\frac14\int d\eta\, d^3x\,\sqrt{-g}\;F_{\mu\nu}^a F^{a\,\mu\nu}, \quad
F_{\mu\nu}^a = \partial_{\mu}A_{\nu}^a - \partial_{\nu}A_{\mu}^a + gf^{abc}A_{\mu}^bA_{\nu}^c .
\end{equation}
The index $a$ labels the generators of the gauge group in the adjoint representation, with $f^{abc}$ the structure constants and $g$ the coupling. In four dimensions the conformal factors cancel: writing~\eqref{eq:ppmetric}
as $g_{\mu\nu} = \eta^{-2}\hat g_{\mu\nu}$, with
$\hat g_{\mu\nu} = {\rm diag}(-1,1,1,1)$, we have $\sqrt{-g} = \eta^{-4}$ and
$F^{a\,\mu\nu} = \eta^4\,\hat g^{\mu\rho}\hat g^{\nu\sigma}F^a_{\rho\sigma}$,
so the action takes the flat-space form
\begin{equation}\label{eq:YMflataction}
S = -\frac14\int d\eta\, d^3x\;\hat g^{\mu\rho}\hat g^{\nu\sigma}
    F_{\mu\nu}^a F^a_{\rho\sigma}.
\end{equation}

\subsection{Free field quantization}\label{sec:quantization}
We work in Coulomb gauge,
\begin{equation}\label{eq:Coulombgauge}
\partial_iA_i^a = 0,
\end{equation}
which leaves $A^a_{\eta}$ unfixed. We also require the fields $A_i^a$ to decay sufficiently
rapidly at spatial infinity. Since $F^a_{\mu\nu}$ is antisymmetric no $\dot{A}^a_{\eta}$
appears in~\eqref{eq:YMflataction}, so $A^a_{\eta}$ carries no conjugate momentum and is an
auxiliary field. Splitting~\eqref{eq:YMflataction} into time and space components with
$F^a_{\eta i} = \dot{A}^a_i - (D_iA_{\eta})^a$, integrating the cross term by parts and
using~\eqref{eq:Coulombgauge},
\begin{equation}\label{eq:31split}
S = \int d\eta\,d^3x\left[\tfrac12\dot{A}^a_i\dot{A}^a_i - \tfrac14 F^a_{ij}F^a_{ij}
    + \tfrac12 A^a_{\eta}\left(-D^2\right)^{ab}A^b_{\eta} + g\,\rho^aA^a_{\eta}\right],
    \qquad \rho^a = f^{abc}A^b_i\dot{A}^c_i ,
\end{equation}
so that varying $A^a_{\eta}$ gives the Gauss constraint
\begin{equation}\label{eq:gauss}
    \left(-\partial^2\right)A^a_{\eta} = -g\,\rho^a + \mathcal{O}(g^2).
\end{equation}

Writing $F^a_{\mu\nu} = G^a_{\mu\nu} + gf^{abc}A^b_{\mu}A^c_{\nu}$, with
$G^a_{\mu\nu} = \partial_{\mu}A^a_{\nu} - \partial_{\nu}A^a_{\mu}$ the abelian field strength, the action separates into a free part and two interaction vertices,
\begin{equation}\label{eq:vertices}
S = -\int d\eta\, d^3x\,\left(\frac14 G_{\mu\nu}^aG^{a\,\mu\nu}
     + V_I^{(3)} + V_I^{(4)}\right),
\end{equation}
\begin{equation}\label{eq:verticesexplicit}
V_I^{(3)} = \tfrac12\,g\,f^{abc}\,G_{\mu\nu}^a\,A^{b\,\mu}A^{c\,\nu} ,
\qquad
V_I^{(4)} = \tfrac14\,g^2 f^{abc}f^{ade}\,A_{\mu}^bA_{\nu}^c A^{d\,\mu}A^{e\,\nu} .
\end{equation}
Since~\eqref{eq:31split} is Gaussian in $A^a_{\eta}$, one may instead complete the square and eliminate it exactly. The
two forms therefore describe the same theory, and each makes something different manifest. Retaining $A^a_{\eta}$ keeps all four components of $A^a_{\mu}$, so the action of an isometry on the fields can be written down directly and Gauss' law~\eqref{eq:gauss} appears as an equation of motion. Eliminating $A_{\eta}$ leaves only the two physical transverse polarizations, at the cost of restricting the vertices to spatial indices and generating in their place an instantaneous, nonlocal Coulomb interaction,
\begin{equation}\label{eq:coulombvertex}
V_I^{\rm C} = \tfrac12\,g^2\,\rho^a\left(-\partial^2\right)^{-1}\rho^a + \mathcal{O}(g^3) ,
\end{equation}
which contributes first at four points. Neither form is $SO(1,4)$ covariant, since the gauge condition~\eqref{eq:Coulombgauge} is not. At three points there is no distinction between the two routes: $V_I^{(3)}$ is already $\mathcal{O}(g)$ and so is evaluated on free fields where $A^a_{\eta}$ vanishes. 

Throughout, $A^a_i$ denotes the interaction-picture field: the solution of the free equation
$\partial_{\mu}G^{a\,\mu\nu} = 0$. 

The free field quantization is that of the Maxwell field on the Poincaré patch, carried out in Coulomb gauge in~\cite{Cotaescu:2008hv} and reviewed in~\cite{Botshekananfard:2026cez}, now with a color index. Expanding in transverse modes of definite helicity $\sigma = \pm\,1$,
\begin{equation}\label{eq:modeexpansion}
A_i^a(\eta, {\bf x}) = \sum_{\sigma} \int \frac{d^3k}{(2\pi)^3} \frac{1}{\sqrt{2k}}\left[\epsilon_{\sigma\,i}({\bf k})e^{-i(k\eta-{\bf k}\cdot{\bf x})}a^a_{\sigma, {\bf k}} + \epsilon^*_{\sigma\,i}({\bf k})e^{i(k\eta-{\bf k}\cdot{\bf x})}a^{\dagger\,a}_{\sigma, {\bf k}}\right],
\end{equation}
with $k \equiv |{\bf k}|$ throughout. The Bunch--Davies vacuum $\ket{0}$ is annihilated by $a^a_{\sigma,{\bf k}}$, 
while $a^{\dagger\,a}_{\sigma,{\bf k}}$ creates a single gluon of color $a$, helicity $\sigma$ and momentum ${\bf k}$.
These obey the commutation relation 
\begin{equation}\label{eq:aadaggcomm}
\left[a^a_{\sigma, {\bf k}},a^{\dagger\,b}_{\sigma', {\bf k}'}\right] = (2\pi)^3\delta^{ab}\delta_{\sigma\sigma'}\delta^{(3)}({\bf k}-{\bf k}').
\end{equation}
The polarization vectors are color independent, and satisfy
\begin{equation}\label{eq:polarizationproperties}
\begin{aligned}
    {\bf k}\cdot\epsilon_{\sigma}({\bf k}) &= 0, &\qquad
    \epsilon_{\sigma}({\bf k})\cdot\epsilon^*_{\sigma'}({\bf k}) &= \delta_{\sigma\sigma'}, \\[4pt]
    {\bf k}\times\epsilon_{\sigma}({\bf k}) &= -i\sigma k\,\epsilon_{\sigma}({\bf k}), &\qquad
    \epsilon_{\sigma}({\bf k})\times\epsilon_{-\sigma}({\bf k}) &= -i\sigma\,\frac{{\bf k}}{k},
\end{aligned}
\end{equation}
together with $\epsilon^*_{\sigma}({\bf k}) = \epsilon_{-\sigma}({\bf k})$ and the
completeness relation
\begin{equation}\label{eq:polarizationcompleteness}
\sum_{\sigma}\epsilon_{\sigma\,i}({\bf k})\,\epsilon^*_{\sigma\,j}({\bf k})
= \delta_{ij} - \frac{k_ik_j}{k^2}.
\end{equation}
\subsection{Conserved charges and the Hamiltonian}\label{sec:charges}
Promoting~\eqref{eq:ppMABs} to operators $Q_{\xi}$ acting on the Fock space, we write
\begin{equation}\label{eq:deltadef}
    \delta_{\xi}\,\mathcal{O} \equiv -i\left[Q_{\xi},\,\mathcal{O}\right]
\end{equation}
for the infinitesimal transformation generated by $Q_{\xi}$, and normalize the charges so
that $\delta_{\xi}$ acts on the fields as a Lie derivative,
\begin{equation}\label{eq:Qfield}
    \delta_{\xi}A_{\mu}^a = \left(\mathcal{L}_{\xi}A^a\right)_{\mu}
    = \xi^{\nu}\partial_{\nu}A_{\mu}^a + \left(\partial_{\mu}\xi^{\nu}\right)A^a_{\nu}.
\end{equation}
The charges obey
\begin{equation}\label{eq:Qalgebra}
    \left[Q_{\xi},\, Q_{\zeta}\right] = -i\,Q_{[\xi,\zeta]},
\end{equation}
with $[\xi,\zeta]$ the Lie bracket of the vector fields, listed in~\eqref{eq:PJDKcomms}. The relation~\eqref{eq:Qfield} holds as written for the generators of ${\rm Sim}(3)$. The
boosts, however, do not preserve Coulomb gauge: acting on a transverse field
$\mathcal{L}_{K_i}A^a$ generates a longitudinal piece, $\partial_j(\mathcal{L}_{K_i}A^a)_j
\neq 0$, so the field leaves the gauge slice~\eqref{eq:Coulombgauge}. The physical
transformation of the field is the Lie derivative followed by a compensating gauge transformation
$\lambda^a_{\xi}$ that restores the gauge choice, i.e.
\begin{equation}\label{eq:compensating}
    \left(\delta_{\xi}A^a\right)_{\mu} = \left(\mathcal{L}_{\xi}A^a\right)_{\mu}
    + \left(D_{\mu}\lambda_{\xi}\right)^a ,
    \qquad \left(D_{\mu}\lambda_{\xi}\right)^a = \partial_{\mu}\lambda_{\xi}^a + gf^{abc}A_{\mu}^b\lambda_{\xi}^c,
\end{equation}
such that
\begin{equation}\label{eq:lambdaconstraint}
    \partial_j\left(\delta_{\xi}A^a\right)_j = 0.
\end{equation}
The compensator is a pure-gauge mode, which is Klein--Gordon orthogonal to every physical mode~\cite{Botshekananfard:2026cez}. Therefore it does not contribute to the free field charges.

The Hermitian charges associated with the generators~\eqref{eq:ppMABs} are listed below.
Throughout the remainder of this section, $\partial_i \equiv \partial/\partial k_i$ denotes a
derivative with respect to momentum. The properties~\eqref{eq:polarizationproperties} fix the
polarization vectors only up to a momentum-dependent phase,
$\epsilon_{\sigma}({\bf k}) \to e^{i\theta_{\sigma}({\bf k})}\epsilon_{\sigma}({\bf k})$, under
which $a^a_{\sigma,{\bf k}} \to e^{-i\theta_{\sigma}({\bf k})}a^a_{\sigma,{\bf k}}$ so that the
field~\eqref{eq:modeexpansion} is unchanged. Momentum derivatives of the mode operators are
therefore taken with the covariant derivative
\begin{equation}\label{eq:Dcov}
    D_i \equiv \partial_i + \alpha_{\sigma,i}({\bf k}),
    \qquad
    \alpha_{\sigma,i}({\bf k}) \equiv \epsilon_{-\sigma,l}({\bf k})\,\partial_i\epsilon_{\sigma,l}({\bf k}) = -\alpha_{-\sigma, i}({\bf k}),
\end{equation}
whose connection shifts as $\alpha_{\sigma,i} \to \alpha_{\sigma,i} + i\partial_i\theta_{\sigma}$,
so that $D_i a^a_{\sigma,{\bf k}}$ transforms like $a^a_{\sigma,{\bf k}}$. The charges are then
\begin{equation}\label{eq:QP}
    Q_{P_i} = \sum_{\sigma}\int\frac{d^3k}{(2\pi)^3}\,a^{\dagger\,a}_{\sigma,{\bf k}}\;
    k_i\;a^{a}_{\sigma,{\bf k}} ,
\end{equation}
\begin{equation}\label{eq:QD}
    Q_{D} = -i\sum_{\sigma}\int\frac{d^3k}{(2\pi)^3}\,a^{\dagger\,a}_{\sigma,{\bf k}}
    \left(k_j\partial_j + \frac32\right)a^{a}_{\sigma,{\bf k}} ,
\end{equation}
\begin{equation}\label{eq:QJ}
    Q_{J_i} = \sum_{\sigma}\int\frac{d^3k}{(2\pi)^3}\,a^{\dagger\,a}_{\sigma,{\bf k}}
    \left(L_i + \sigma\,\frac{k_i}{k}\right)a^{a}_{\sigma,{\bf k}} ,\qquad L_i \equiv -i\left({\bf k}\times {\bf D}\right)_i, 
\end{equation}
\begin{equation}\label{eq:QK}
    Q_{K_i} = \sum_{\sigma}\int\frac{d^3k}{(2\pi)^3}\,a^{\dagger\,a}_{\sigma,{\bf k}}
    \left(\mathcal{K}_i - \frac{2\sigma}{k}L_i\right)a^{a}_{\sigma,{\bf k}}.
\end{equation}
Here the color index is summed, and the operator appearing in the boost charge is
\begin{equation}\label{eq:boostpiece}
\mathcal{K}_i = -k_iD_jD_j + 2k_jD_jD_i + 3D_i - \frac{5k_i}{4k^2}.
\end{equation}

The free Hamiltonian is the charge associated with the proper time
generator~\eqref{eq:DaP} for a given eternal observer,
\begin{equation}\label{eq:freeH0}
H_0({\bf a}) = Q_{D} + {\bf a}\cdot Q_{{\bf P}},
\end{equation}
which annihilates the Bunch--Davies vacuum: every eternal observer assigns $\ket{0}$ zero
energy. 

The action on the mode operators follows directly:
\begin{equation}\label{eq:Qacomms}
\begin{aligned}
\left[Q_{P_i},\, a^a_{\sigma,{\bf k}}\right] &= -k_i\,a^a_{\sigma,{\bf k}},
&&\quad \left[Q_{P_i},\, a^{\dagger\,a}_{\sigma,{\bf k}}\right] = k_i\,a^{\dagger\,a}_{\sigma,{\bf k}}, \\[4pt]
\left[Q_{D},\, a^a_{\sigma,{\bf k}}\right] &= i\left(k_j\partial_j + \tfrac32\right)a^a_{\sigma,{\bf k}},
&&\quad \left[Q_{D},\, a^{\dagger\,a}_{\sigma,{\bf k}}\right] = i\left(k_j\partial_j + \tfrac32\right)a^{\dagger\,a}_{\sigma,{\bf k}}, \\[4pt]
\left[Q_{J_i},\, a^a_{\sigma,{\bf k}}\right] &= -\left(L_i + \sigma\tfrac{k_i}{k}\right)a^a_{\sigma,{\bf k}},
&&\quad \left[Q_{J_i},\, a^{\dagger\,a}_{\sigma,{\bf k}}\right] = \left(\bar{L}_i + \sigma\tfrac{k_i}{k}\right)a^{\dagger\,a}_{\sigma,{\bf k}}, \\[4pt]
\left[Q_{K_i},\, a^a_{\sigma,{\bf k}}\right] &= -\left(\mathcal{K}_i - \tfrac{2\sigma}{k}L_i\right)a^a_{\sigma,{\bf k}},
&&\quad \left[Q_{K_i},\, a^{\dagger\,a}_{\sigma,{\bf k}}\right]=\left(\bar{\mathcal{K}}_i - \tfrac{2\sigma}{k}\bar{L}_i\right)a^{\dagger\,a}_{\sigma,{\bf k}},
\end{aligned}
\end{equation}
where $\bar{X}$ denotes complex conjugation of the operator $X$, so that
$\bar{D}_i = \partial_i + \alpha_{-\sigma,i}$, $\bar{L}_i = i({\bf k}\times\bar{\bf D})_i$,
and $\bar{\mathcal{K}}_i$ is~\eqref{eq:boostpiece} with $D_i \to \bar{D}_i$. Note that
$[Q_D, Q_{P_i}] \neq 0$: the states created by $a^{\dagger\,a}_{\sigma,{\bf k}}$ carry
definite momentum and helicity but not definite energy. Correspondingly, $Q_D$ acts on the
mode operators as a dilation in momentum space, so the free
evolution generated by $H_0$ rescales ${\bf k}$ rather than leaving it fixed. 

The charges~\eqref{eq:QP}--\eqref{eq:QK} were obtained by acting with each generator on the
free field and reading off the action on the mode functions. They may equivalently be
obtained from the conserved Noether current $J^{\mu}_{\xi} = -T^{\mu}{}_{\nu}\xi^{\nu}$,
\begin{equation}\label{eq:stresstensor}
T_{\mu\nu} = F^a_{\mu\rho}F^a_{\nu\sigma}g^{\rho\sigma}
           - \tfrac14 g_{\mu\nu}F^a_{\rho\sigma}F^{a\,\rho\sigma},
\end{equation}
integrated over a Cauchy slice of constant conformal time,
\begin{equation}\label{eq:Qfromstress}
Q_{\xi} = -\int_{\Sigma_{\eta}} d\Sigma_{\mu}\;T^{\mu}{}_{\nu}\,\xi^{\nu}
        = \int d^3x\;\frac{1}{\eta^4}\,T^{\eta}{}_{\nu}\,\xi^{\nu},
\end{equation}
with the free charges coming from the free part of $T_{\mu\nu}$. This second route is the
one we need for the interaction Hamiltonian, which is found by contracting the conserved current of~\eqref{eq:DaP} with the interacting part
of~\eqref{eq:stresstensor},
\begin{equation}\label{eq:HI}
H_I = \int d^3x\;\frac{1}{\eta^4}\left(-\eta\,T^{I\,\eta}{}_{\eta}
      - (x_i-a_i)\,T^{I\,\eta}{}_i\right)
    = \int d^3x\;\left(V_I^{(3)} + V_I^{(4)}\right),
\end{equation}
evaluated on the slice $\eta = -1$, corresponding to $\tau = 0$. The contribution from
$T^{I\,\eta}{}_i$ integrates to zero using repeated integration by parts
and~\eqref{eq:Coulombgauge}; since ${\bf a}$ appears only in that term, $H_I$ is independent
of the observer. Together these give the full Hamiltonian
\begin{equation}\label{eq:fullH}
H({\bf a}) = H_0({\bf a}) + H_I .
\end{equation}

\section{A Poincar\'e patch S-matrix}\label{sec:Smatrix}

The Killing vector~\eqref{eq:DaP} associated with the free Hamiltonian is not timelike
everywhere in the patch. Its norm is
\begin{equation}\label{eq:freehamnorm}
\frac{1}{\eta^2}\left(-\eta^2 + |{\bf x}-{\bf a}|^2\right),
\end{equation}
so it is timelike only for $|{\bf x}-{\bf a}| < -\eta$, the static patch of the observer at
${\bf a}$, and spacelike beyond her horizon. This does not obstruct the construction below,
which uses two properties of~\eqref{eq:freeH0} that hold globally. First, it is a Noether
charge, obtained from~\eqref{eq:Qfromstress} by integrating a conserved current over a full
Cauchy slice, and is therefore self-adjoint regardless of the causal character
of~\eqref{eq:DaP} at any given point. Second, the flow it generates is an automorphism of the
whole patch, taking slices of constant $\eta$ to slices of constant $\eta$, so that ordering
along that flow is ordering with respect to a spacelike foliation of the entire region. The same distinction is familiar in Minkowski space, where the boost generator is self-adjoint and
exponentiates to a unitary flow on the full Hilbert space, although the boost Killing vector is
future-directed timelike only inside the Rindler wedge~\cite{Bisognano:1975ih,Bisognano:1976za}.

Following~\cite{Taylor:2024vdc}, we construct the S-matrix using the time-independent
Hamiltonian $H({\bf a})$ of~\eqref{eq:fullH}. It is defined as the overlap of asymptotic in-
and out-states
\begin{equation}\label{eq:Sdef1}
    S_{n \to m} = {}_{\rm out}\!\braket{m|n}_{\rm in}.
\end{equation}
We assume that interactions may be neglected at asymptotically early and late proper times
along the worldline of an eternal observer,\footnote{\label{fn:adiabatic}This is the adiabatic hypothesis. Yang--Mills
is classically conformal, so in conformal time the dynamics is that of flat
space throughout the patch, and the two limits differ only through
$d\eta'=-\eta'\,d\tau$. By~\eqref{eq:H0fieldflow}, a vertex at proper time
$\tau$ sits at $\eta'=-e^{-\tau}$ and carries a factor $-\eta'=e^{-\tau}$,
so at fixed comoving momenta $H_I(\tau)\sim e^{-\tau}$ and
$\int^{\tau_+}\!d\tau\,H_I(\tau)$ converges as $\tau_+\to+\infty$. As
$\tau_-\to-\infty$ the factor grows instead, and the limit is defined, as in
flat space, by an $i\epsilon$ prescription: the Bunch--Davies rotation
$\eta\to\eta-i\epsilon$ damps the far past.}
so that the exact states $\ket{n}_{\rm in}$ and $\ket{m}_{\rm out}$ are described in those
limits by free-particle states $\ket{n}$ and $\ket{m}$ built from the creation operators
of~\eqref{eq:modeexpansion}. Both are constructed at the common time $\tau = 0$, where the
overlap in~\eqref{eq:Sdef1} is evaluated: a free state is evolved to $\tau_{\mp}$ with
$H_0({\bf a})$ and returned to $\tau = 0$ with the full $H({\bf a})$,
\begin{equation}\label{eq:inoutdef}
    \ket{n}_{\rm in} = \lim_{\tau_-\to-\infty} e^{iH({\bf a})\tau_-}e^{-iH_0({\bf a})\tau_-}\ket{n},
    \qquad
    \ket{m}_{\rm out} = \lim_{\tau_+\to+\infty} e^{iH({\bf a})\tau_+}e^{-iH_0({\bf a})\tau_+}\ket{m}.
\end{equation}
The two evolutions cancel except for the interaction, and under the
adiabatic hypothesis (footnote~\ref{fn:adiabatic}) the accumulated
interaction is finite at both ends, so the limits exist. We may therefore write
$S_{n\to m}$ as
\begin{equation}\label{eq:Sdef2}
    S_{n\to m} = \bra{m}\,U(\tau_+,\tau_-)\,\ket{n},
    \qquad
    U(\tau_+,\tau_-) = e^{iH_0({\bf a})\tau_+}\,e^{-iH({\bf a})(\tau_+-\tau_-)}\,e^{-iH_0({\bf a})\tau_-},
\end{equation}
evaluated as $\tau_{\pm}\to\pm\infty$. In this limit $U$ reduces to the Dyson series
\begin{equation}\label{eq:Dyson}
    U(+\infty,-\infty) = T\exp\left(-i\int_{-\infty}^{\infty}\!d\tau\;H_I(\tau)\right),
    \qquad
    H_I(\tau) = e^{iH_0({\bf a})\tau}H_I\,e^{-iH_0({\bf a})\tau},
\end{equation}
with $H_I$ given by~\eqref{eq:HI}. Conjugation by $e^{iH_0({\bf a})\tau}$ implements the
isometry generated by $H_0({\bf a})$,
\begin{equation}\label{eq:H0coordflow}
    \eta \to \eta' = e^{-\tau}\eta, \qquad
    {\bf x} \to {\bf x}' = {\bf a} + e^{-\tau}({\bf x}-{\bf a}),
\end{equation}
\begin{equation}\label{eq:H0fieldflow}
    e^{iH_0({\bf a})\tau}A^a_i(\eta,{\bf x})\,e^{-iH_0({\bf a})\tau}
    = \frac{\partial x'^j}{\partial x^i}A^a_j(\eta',{\bf x}')
    = e^{-\tau}A^a_i(\eta',{\bf x}').
\end{equation}
Each vertex in~\eqref{eq:verticesexplicit} therefore transforms as
$V_I \to e^{-4\tau}V_I(\eta',{\bf x}')$. Together with $d^3x = e^{3\tau}d^3x'$ this gives
$H_I(\tau) = -\eta'\!\int d^3x'\,V_I(\eta',{\bf x}')$, and since $d\eta' = -\eta'\,d\tau$ the
$\tau$-integral assembles into an integral over the patch,
\begin{equation}\label{eq:tautopatch}
    \int_{-\infty}^{\infty}\! d\tau\;H_I(\tau)
    = \int_{-\infty}^{0}\! d\eta \int\! d^3x\; :\!V_I^{(3)} + V_I^{(4)}\!: \, .
\end{equation}
The infinite proper time of the observer has been compressed into the finite conformal range
$\eta\in(-\infty,0)$, and we arrive at
\begin{equation}\label{eq:Sfinal}
    S_{n\to m} = \bra{m}\,T\exp\left(-i\int_{-\infty}^{0}\! d\eta \int\! d^3x\;
    :\!V_I^{(3)} + V_I^{(4)}\!:\right)\ket{n} .
\end{equation}

This construction was described relative to a particular eternal observer, but the result
does not depend on which. The flow~\eqref{eq:H0coordflow} acts on conformal time as
$\eta\to e^{-\tau}\eta$ for every ${\bf a}$, so all eternal observers induce the same
foliation and the same time-ordering in~\eqref{eq:Dyson}, and the interaction
Hamiltonian~\eqref{eq:HI} is itself independent of ${\bf a}$. The operator
in~\eqref{eq:Sfinal} is therefore the same for every eternal observer, and only the
definitions~\eqref{eq:inoutdef} of the asymptotic states retain any dependence.

That residual dependence is a phase. Eternal observers are related by translations, so their
Hamiltonians are related by conjugation,
$H_0({\bf a}+{\bf c}) = T_{\bf c}H_0({\bf a})T_{\bf c}^{-1}$ with
$T_{\bf c} = e^{-i{\bf c}\cdot Q_{\bf P}}$, and~\eqref{eq:inoutdef} then gives
$S^{({\bf a}+{\bf c})} = T_{\bf c}\,S^{({\bf a})}\,T_{\bf c}^{-1}$. Independence of ${\bf a}$
is therefore the statement $[Q_{P_i},S] = 0$, which in the momentum basis is momentum
conservation,
\begin{equation}\label{eq:momcons}
    S_{n\to m} = (2\pi)^3\,\delta^{(3)}\!\left({\bf k}_{\rm tot}\right)\,\tilde{S}_{n\to m},
    \qquad {\bf k}_{\rm tot} = \sum_b {\bf k}_b ,
\end{equation}
where all momenta are taken incoming, outgoing legs carrying ${\bf k}_b \to -{\bf k}_b$.
Conjugation by $T_{\bf c}$ multiplies a momentum-basis matrix element by
$e^{i{\bf c}\cdot{\bf k}_{\rm tot}}$, which is unity on the support of the delta function
in~\eqref{eq:momcons}, so the reduced amplitude $\tilde{S}_{n\to m}$ is the same for every
eternal observer, and nothing distinguishes one member of the congruence from another.

What remains is the choice of congruence itself, and it is not a choice. An S-matrix requires
worldlines with an asymptotic past and an asymptotic future in the region where the theory is
defined. By~\eqref{eq:etapnot0} every timelike geodesic with ${\bf p}\neq0$ crosses the past
horizon at the finite proper time $\tau^*$, entering the patch from outside and supporting no
in-state, so completeness to the past forces ${\bf p}=0$ on its own; and
by~\eqref{eq:etap0} the eternal worldlines ${\bf x} = {\bf a}$ are then the only geodesics
that survive. They are the unique complete geodesic congruence available in the expanding
patch.

Since
$\mathcal{I}^+$ is spacelike, no single observer can measure a complete set
of out-states: every mode of fixed comoving momentum is stretched beyond her
horizon, $|{\bf x}-{\bf a}| = -\eta$, as $\tau\to+\infty$. Like late-time
cosmological correlators, the S-matrix is therefore a
meta-observable~\cite{Witten:2001kn} rather than an operational one. Its content lies in the symmetry constraints it satisfies
(Section~\ref{sec:WI}) and in its flat limit (Section~\ref{sec:flatlimit}),
which in some channels describes sub-Hubble scattering that a single
observer can carry out.

The construction of this S-matrix is closest to the Bunch--Davies S-matrix of~\cite{Melville:2023kgd}, which is also formulated in the expanding Poincar\'e patch and also builds both the in- and the out-states from Bunch--Davies Fock states. However, the two routes to the definition are different: \cite{Melville:2023kgd} obtain theirs by an LSZ reduction of time-ordered correlators, whereas ours follows from Hamiltonian evolution in the proper time of an eternal observer. What the Hamiltonian route makes explicit is the Killing vector whose proper time the limits in~\eqref{eq:inoutdef} refer to. 

\subsection{3-pt amplitudes}\label{sec:amplitudes}
We now evaluate~\eqref{eq:Sfinal} at three points. All momenta are taken incoming, as
in~\eqref{eq:momcons}, and the magnitudes $k_b$ are treated as independent variables, equal
to $+|{\bf k}_b|$ for an incoming leg and $-|{\bf k}_b|$ for an outgoing one. We write
\begin{equation}\label{eq:kT}
    k_T \equiv \sum_b k_b
\end{equation}
for the total ``energy''\footnote{We use the term `energy' here borrowed from flat space;
$k$ is not an eigenvalue of $H_0$, and by $[Q_D,Q_{P_i}]\neq0$ these states have no
definite energy at all. We keep the name because the $k_T = 0$ singularity is standard
terminology and is where the flat-space amplitude is recovered.} flowing into the vertex.
With this convention every external leg contributes the same factor $e^{-ik_b\eta}$
to~\eqref{eq:Sfinal}: an incoming leg contracts with the annihilation part
of~\eqref{eq:modeexpansion} and supplies $e^{-i|{\bf k}_b|\eta}$, an outgoing leg contracts
with the creation part and supplies $e^{+i|{\bf k}_b|\eta}$, and the two coincide once
$k_b = \mp|{\bf k}_b|$. A leg is moved between the in- and out-states by
\begin{equation}\label{eq:crossing}
    \left({\bf k}_b,\,k_b,\,\sigma_b\right) \;\longrightarrow\;
    \left(-{\bf k}_b,\,-k_b,\,-\sigma_b\right) ,
\end{equation}
the helicity flipping because the polarization vectors depend only on the direction
$\hat{\bf k}_b$ and satisfy $\epsilon_{\sigma}(-\hat{\bf k}) = \epsilon_{-\sigma}(\hat{\bf k})$.

As in flat space, the lower endpoint of the time integral~\eqref{eq:Sfinal} is regulated by
the Bunch-Davies $i\epsilon$ prescription
(see footnote~\ref{fn:adiabatic}). With all legs incoming,
\begin{equation}\label{eq:etaint}
    \int_{-\infty}^{0}\! d\eta\; e^{-ik_T(\eta-i\epsilon)} = \frac{i}{k_T + i\epsilon},
\end{equation}
where the $i\epsilon$ matters only at $k_T=0$; we drop it below and restore
it on that locus. Since every channel is obtained from the all-incoming one by~\eqref{eq:crossing}, we write
$\tilde S_n \equiv \tilde S_{n\to0}$ for the reduced amplitude with all $n$ legs incoming, and
obtain the amplitude for any other channel by crossing its outgoing legs. The three gluon S-matrix element is given by
\begin{equation}\label{eq:S3}
\tilde{S}_{3} = i\,g\,f^{a_1a_2a_3}\,\frac{1}{2^{3/2}\sqrt{k_1k_2k_3}}\;
\frac{1}{k_T}\;
\left[\left(({\bf k}_1 - {\bf k}_3)\cdot\epsilon_2\right)
\left(\epsilon_1\cdot\epsilon_3\right)
+ {\rm cyclic}\right],
\end{equation}
where $\epsilon_b \equiv \epsilon_{\sigma_b}({\bf k}_b)$. As a quick example, applying~\eqref{eq:crossing} to
$\tilde{S}_{3}$ with leg $3$ now taken to be outgoing gives
\begin{equation}\label{eq:s2to1}
\begin{aligned}
\tilde{S}_{2 \to 1}  = g\,f^{a_1a_2a_3}\,\frac{1}{2^{3/2}\sqrt{k_1k_2k_3}}\frac{1}{k_T}
\left[(({\bf k}_1-(-{\bf k}_3))\cdot\epsilon_2)(\epsilon_1\cdot\epsilon^*_3) + {\rm cyclic}\right],
\end{aligned}
\end{equation}
where $k_T = k_1+k_2-k_3$. 

Since the mode functions~\eqref{eq:modeexpansion} are the flat ones, the time integral we performed in~\eqref{eq:etaint} is the same one that computes the late-time wavefunction of the universe for gauge fields. It agrees structurally with the three-point function of~\cite{Maldacena:2011nz}, differing only in the normalization of the external legs:
\begin{equation}\label{eq:Stopsi}
\tilde{S}_{3} \;=\; \left[\prod_{b=1}^{3}\frac{1}{\sqrt{2k_b}}\right]\psi_3 ,
\end{equation}
which is the general relation between the de Sitter S-matrix and the wavefunction established in~\cite{Melville:2023kgd}. The wavefunction coefficients are defined with all legs incoming; sending $(\mathbf{k}_b, k_b,\sigma_b)$ to
$(-\mathbf{k}_b,-k_b,-\sigma_b)$ produces distinct $n\to m$ processes, and it is
this continuation, rather than the value of the integral, that distinguishes the S-matrix from the wavefunction coefficients.

The three-point amplitude~\eqref{eq:S3} takes a compact form in spinor-helicity
variables~\cite{Maldacena:2011nz}, whose definitions and conventions are collected in
Appendix~\ref{app:spinorhelicity}. Using three-momentum conservation, in the form of the
identities~\eqref{eq:threepointidentities}, the independent helicity configurations become
\begin{equation}\label{eq:S3helicity}
\begin{aligned}
\tilde{S}^{+++}_{3} &= \phantom{-}\frac{i\,g\,f^{a_1a_2a_3}}{2\,(k_1k_2k_3)^{3/2}}\;
\langle\bar{3}\bar{2}\rangle\langle\bar{1}\bar{2}\rangle\langle\bar{1}\bar{3}\rangle ,
\\[6pt]
\tilde{S}^{++-}_{3} &= -\frac{i\,g\,f^{a_1a_2a_3}}{8\,(k_1k_2k_3)^{3/2}}\;
   \frac{\prod_b\left(k_T-2k_b\right)}{k_T}\;
   \frac{\langle\bar{1}\bar{2}\rangle^{3}}
        {\langle\bar{1}\bar{3}\rangle\langle\bar{2}\bar{3}\rangle} ,
\\[6pt]
\tilde{S}^{+--}_{3} &= -\frac{i\,g\,f^{a_1a_2a_3}}{8\,(k_1k_2k_3)^{3/2}}\;
   \frac{\prod_b\left(k_T-2k_b\right)}{k_T}\;
   \frac{\langle23\rangle^{3}}{\langle21\rangle\langle31\rangle} ,
\\[6pt]
\tilde{S}^{---}_{3} &= -\frac{i\,g\,f^{a_1a_2a_3}}{2\,(k_1k_2k_3)^{3/2}}\;
   \langle32\rangle\langle12\rangle\langle13\rangle .
\end{aligned}
\end{equation}

\subsection{4-pt amplitudes}
At four points three structures contribute at $\mathcal{O}(g^2)$: the quartic vertex
$V_I^{(4)}$, the instantaneous Coulomb vertex $V_I^{\rm C}$, and the exchange of an
internal gluon between two cubic vertices. For the exchange diagram we will need the
propagator of the transverse field, which is the flat Feynman
propagator in Coulomb gauge,
\begin{equation}\label{eq:feynprop}
G^{ab}_{mn}\!\left({\bf P};\eta_1,\eta_2\right)
 = \delta^{ab}\,\Pi_{mn}({\bf P})\,
   \frac{e^{-i\left(P-i\epsilon\right)\left|\eta_1-\eta_2\right|}}{2P},
\qquad
\Pi_{mn}({\bf P}) = \delta_{mn} - \frac{P_mP_n}{P^2},
\end{equation}
with $P = |{\bf P}|$.
Writing each contribution as $N_4$, $N_{\rm C}$ and $N_3$, respectively, the four-gluon S-matrix element is
\begin{equation}\label{eq:S4}
\tilde{S}_{4} = \frac{g^2}{4}\,\frac{1}{k_T\sqrt{k_1k_2k_3k_4}}
\left(N_4 - N_{\rm C} - N_3\right),
\end{equation}
where
\begin{equation}\label{eq:N4}
N_4 = c_s\Big[\left(\epsilon_1\!\cdot\!\epsilon_3\right)\left(\epsilon_2\!\cdot\!\epsilon_4\right)
    - \left(\epsilon_1\!\cdot\!\epsilon_4\right)\left(\epsilon_2\!\cdot\!\epsilon_3\right)\Big]
    \;+\; c_t\Big[\left(2\leftrightarrow3\right)\Big]
    \;-\; c_u\Big[\left(2\leftrightarrow4\right)\Big],
\end{equation}
\begin{equation}\label{eq:NC}
N_{\rm C} = c_s\left[\frac{\left(k_1-k_2\right)\left(k_3-k_4\right)}{P_s^2}\,
    \left(\epsilon_1\!\cdot\!\epsilon_2\right)\left(\epsilon_3\!\cdot\!\epsilon_4\right)\right]
    \;+\; c_t\Big[\left(2\leftrightarrow3\right)\Big]
    \;-\; c_u\Big[\left(2\leftrightarrow4\right)\Big],
\end{equation}
\begin{equation}\label{eq:N3}
\begin{aligned}
N_3 & = c_s\left[\frac{1}{2P_s}\left(\frac{1}{k_1+k_2-P_s}+\frac{1}{k_3+k_4-P_s}\right)
    \mathcal{V}_m(1,2)\,\Pi_{mn}\!\left({\bf P}_s\right)\,\mathcal{V}_n(3,4)\right]
    \\ 
    & \quad +\; c_t\Big[\left(2\leftrightarrow3\right)\Big]
    \;-\; c_u\Big[\left(2\leftrightarrow4\right)\Big].
\end{aligned}
\end{equation}
Each term is a sum over the $s$-, $t$-, and $u$-channels, labeled by the color factors and internal momenta
\begin{equation}\label{eq:channels}
\begin{aligned}
c_s & = f^{a_1a_2\ell}f^{a_3a_4\ell}, \qquad
{\bf P}_s = {\bf k}_1+{\bf k}_2,\\
c_t & = f^{a_1a_3\ell}f^{a_2a_4\ell}, \qquad
{\bf P}_t = {\bf k}_1+{\bf k}_3,\\
c_u & = f^{a_1a_4\ell}f^{a_2a_3\ell}, \qquad
{\bf P}_u = {\bf k}_1+{\bf k}_4.
\end{aligned}
\end{equation}
The color factors obey the relation
\begin{equation}\label{eq:colorjacobi}
c_s - c_t + c_u = 0,
\end{equation}
which follows from the Jacobi identity for the structure constants. The color factor of each
term is displayed explicitly, and $(2\leftrightarrow3)$, $(2\leftrightarrow4)$ instruct that
the leg labels be exchanged in the bracketed tensor structure only, with
${\bf P}_s\to{\bf P}_t$ and ${\bf P}_s\to{\bf P}_u$ respectively where an internal momentum
appears. Here $\Pi_{mn}$ is the transverse projector onto the internal line and $\mathcal{V}_m(r,s)$
is the cubic vertex with the internal leg left open,
\begin{equation}\label{eq:Vhalf}
\mathcal{V}_m(r,s) = 2\left({\bf k}_r\!\cdot\!\epsilon_s\right)\epsilon_{r\,m}
    - 2\left({\bf k}_s\!\cdot\!\epsilon_r\right)\epsilon_{s\,m}
    - \left({\bf k}_r-{\bf k}_s\right)_m\left(\epsilon_r\!\cdot\!\epsilon_s\right),
\end{equation}
the momentum of the internal line having been eliminated using
${\bf k}_r\!\cdot\!\epsilon_r = 0$, so that $\mathcal{V}_m(r,s)$ depends only on the two
external legs at that vertex. 
It is convenient to trade the color factors~\eqref{eq:channels} for single traces of generators\footnote{We use the convention $\Tr(T^aT^b) = \delta^{ab}$ so that $f^{abc} = -i/\sqrt{2}\Tr([T^a, T^b]T^c)$.}, so that
\begin{equation}\label{eq:S4color}
\tilde{S}_{4} = -\frac{g^2}{8}\,\frac{1}{k_T\sqrt{k_1k_2k_3k_4}}
\sum_{\sigma\in S_3}
{\rm Tr}\!\left(T^{a_1}T^{a_{\sigma(2)}}T^{a_{\sigma(3)}}T^{a_{\sigma(4)}}\right)
A\!\left(1,\sigma(2),\sigma(3),\sigma(4)\right),
\end{equation}
with the sum over permutations of the external leg labels $\{2,3,4\}$. The partial amplitudes obey the reflection and decoupling identities
\begin{equation}
    A(1,2,3,4) = A(1,4,3,2), \qquad A(1,2,3,4)+A(1,3,4,2)+A(1,4,2,3) = 0,
\end{equation}
so only two of the six
terms are independent. For the ordering $(1,2,3,4)$,
\begin{equation}\label{eq:A1234}
\begin{aligned}
A(1,2,3,4) & = 2(\epsilon_1\!\cdot\!\epsilon_3)(\epsilon_2\!\cdot\!\epsilon_4)
    - (\epsilon_1\!\cdot\!\epsilon_4)(\epsilon_2\!\cdot\!\epsilon_3)
    - (\epsilon_1\!\cdot\!\epsilon_2)(\epsilon_3\!\cdot\!\epsilon_4) \\[4pt]
& \quad - \frac{(k_1-k_2)(k_3-k_4)}{P_s^2}\,
    (\epsilon_1\!\cdot\!\epsilon_2)(\epsilon_3\!\cdot\!\epsilon_4)
  + \frac{(k_1-k_4)(k_2-k_3)}{P_u^2}\,
    (\epsilon_1\!\cdot\!\epsilon_4)(\epsilon_2\!\cdot\!\epsilon_3) \\[4pt]
& \quad - \frac{1}{2P_s}\left(\frac{1}{k_1+k_2-P_s} + \frac{1}{k_3+k_4-P_s}\right)
    \mathcal{V}_m(1,2)\,\Pi_{mn}({\bf P}_s)\,\mathcal{V}_n(3,4) \\[4pt]
& \quad + \frac{1}{2P_u}\left(\frac{1}{k_1+k_4-P_u} + \frac{1}{k_2+k_3-P_u}\right)
    \mathcal{V}_m(1,4)\,\Pi_{mn}({\bf P}_u)\,\mathcal{V}_n(2,3).
\end{aligned}
\end{equation}

Similar to the three-point case, the contact contributions match the
corresponding wavefunction coefficient~\cite{Baumann:2020dch} up to the normalization of the
external legs. The exchange
contributions differ, however, since the wavefunction is computed with the bulk-to-bulk propagator,
\begin{equation}\label{eq:bulk2bulk}
\mathcal{G}_{mn}\!\left({\bf P};\eta_1,\eta_2\right)
 = \Pi_{mn}({\bf P})\,\frac{1}{2P}\left[
   e^{-iP\left|\eta_1-\eta_2\right|} - e^{iP\left(\eta_1+\eta_2\right)}\right],
\end{equation}
which differs from~\eqref{eq:feynprop} by a homogeneous term chosen so that it vanishes at
$\mathcal{I}^+$. That term factorizes in
$\eta_1$ and $\eta_2$, so the exchange integral becomes a product of partial energies rather
than a sum,
\begin{equation}\label{eq:bbint}
\begin{aligned}
&-\frac{1}{2P_s\,k_T}\left[\frac{1}{k_1+k_2-P_s} + \frac{1}{k_3+k_4-P_s}\right]
 + \frac{1}{2P_s\left(k_1+k_2-P_s\right)\left(k_3+k_4-P_s\right)}\\
& \qquad  = \frac{1}{k_T\left(k_1+k_2-P_s\right)\left(k_3+k_4-P_s\right)}.
\end{aligned}
\end{equation}
The two prescriptions therefore share both the total-energy pole and the same partial
energies.

\section{Ward identities}\label{sec:WI}
A generator $\xi$ is realized as a symmetry of the evolution operator,
\begin{equation}\label{eq:QUcomm}
    \left[Q_{\xi},\,U\right] = 0 ,
\end{equation}
when its flow maps the interaction density into a total derivative,
preserves the region of integration in~\eqref{eq:tautopatch}, and
preserves the gauge. Together
with $Q_{\xi}\ket{0} = 0$, this gives $[Q_{\xi},S] = 0$, and sandwiching between free states
and using~\eqref{eq:Qacomms} turns it into a homogeneous constraint equation for
$\tilde{S}_{n\to m}$ in the external momenta and helicities.

For the generators of ${\rm Sim}(3)$, \eqref{eq:QUcomm} holds. They preserve the patch, so
the domain of~\eqref{eq:tautopatch} is unchanged, and they preserve the gauge
slice~\eqref{eq:Coulombgauge} so the fields transform by the Lie derivative alone. The
resulting identities are the familiar ones and are already built into~\eqref{eq:S3}:
translations give momentum conservation~\eqref{eq:momcons}, rotations require $\tilde{S}$ to
be a rotational invariant carrying the appropriate helicity weights, and the dilation fixes
its overall scaling weight in the momenta.

For the boosts, \eqref{eq:QUcomm} fails, and the obstruction is the gauge slice. $K_i$ does not preserve~\eqref{eq:Coulombgauge}, so the fields transform by the Lie derivative
together with a compensating gauge transformation~\eqref{eq:compensating}. The domain
of~\eqref{eq:tautopatch} is not an obstruction, even though the boosts do not map the patch
to itself; the variation of the interaction density is a total
derivative which vanishes upon integration.

Recall that the compensator $\lambda_{\xi}^a$ is the solution of
\begin{equation}\label{eq:lambdadiffeq}
\partial_j(\delta_{\xi}A^a)_j = 0, \qquad (\delta_{\xi}A^a)_j = (\partial_{\xi}A^a)_{j} + \partial_j\lambda_{\xi}^a + gf^{abc}A_{j}^b\lambda_{\xi}^c.
\end{equation}
We will denote by $\lambda_i^a$ the compensator associated with the boost $K_i$. In general, $\lambda_i^a$ can be expanded as
\begin{equation}
\lambda_i^{a} = \lambda_i^{a\,(0)} + g\lambda_i^{a\,(1)} + \mathcal{O}(g^2).
\end{equation}
Solving~\eqref{eq:lambdadiffeq} order by order, we find the leading term to be
\begin{equation}\label{eq:lambda0def}
\lambda_i^{a\,(0)}(\eta, {\bf x}) = \sum_{\sigma}\int \frac{d^3k}{(2\pi)^3}\left[\epsilon_{\sigma,\,i}(\hat{\bf k})\frac{2(1+ik\eta)}{k^2\sqrt{2k}}e^{-ik\eta + i{\bf k}\cdot{\bf x}}\,a^a_{\sigma, {\bf k}} + {\rm h.c.}\right].
\end{equation}
The origin of the inhomogeneous term in the three-point boost Ward identity is the
$\mathcal{O}(g)$ piece of $\left[Q_{K_i},U\right]$. Only the cubic vertex contributes at this
order, and its Lie derivative is a total derivative,
$\mathcal{L}_{K_i}V_I^{(3)} = \partial_{\mu}(K_i^{\mu}V_I^{(3)})$, which integrates to zero:
the spatial part vanishes by the fall-off conditions of Section~\ref{sec:quantization}, and the
temporal part vanishes at both endpoints, at $\eta\to0^-$ because $K_i^{\eta} = -2\eta x_i$
and at $\eta\to-\infty$ by the $i\epsilon$. The remaining term is the contribution from the
compensator, which gives $gf^{abc}G^{a\,\mu\nu}(\partial_{\mu}\lambda^{b\,(0)}_i)A^c_{\nu}$.
Integrating by parts and using the free equations of motion
$\partial_{\mu}G^{a\,\mu\nu} = 0$ turns this into a total derivative, whose spatial part
vanishes, leaving
\begin{equation}\label{eq:QUboundary}
\left[Q_{K_i},U\right]
= gf^{abc}\left[\int d^3x\;\dot{A}^{a\,\nu}\,\lambda^{b\,(0)}_i\,A^c_{\nu}\right]^{\eta\to0^-}_{\eta\to-\infty-i\epsilon} + \mathcal{O}(g^2) .
\end{equation}
The lower endpoint vanishes under the Bunch--Davies $i\epsilon$ prescription, so this is
entirely a boundary term at $\mathcal{I}^+$. The boost Ward identity for the three point
S-matrix element~\eqref{eq:S3} is then
\begin{equation}
\bra{0}U\left[Q_{K_i}, a^{\dagger a_1}_{\sigma_1, {\bf k}_1}a^{\dagger a_2}_{\sigma_2, {\bf k}_2}a^{\dagger a_3}_{\sigma_3, {\bf k}_3}\right]\ket{0} = -\bra{0}\left[Q_{K_i}, U\right]a^{\dagger a_1}_{\sigma_1, {\bf k}_1}a^{\dagger a_2}_{\sigma_2, {\bf k}_2}a^{\dagger a_3}_{\sigma_3, {\bf k}_3}\ket{0},
\end{equation}
which, using~\eqref{eq:Qacomms} and the boundary term above, evaluates to
\begin{equation}\label{eq:S3WI}
\sum_{b=1}^3\left(\bar{\mathcal{K}}_i^b - \frac{2\sigma_b}{k_b}\bar{L}^b_i\right)\tilde{S}_{3}\left(\{{\bf k}_b, \sigma_b, a_b\}\right) = -ig \frac{f^{a_1a_2a_3}}{\sqrt{2k_1k_2k_3}}\Big[\epsilon_{2,i}\left(\epsilon_{1}\cdot\epsilon_{3}\right)\frac{k_1-k_3}{k_2^2} + {\rm cyc.}\Big],
\end{equation}
where it is understood that $\bar{\mathcal{K}}_i^b - \frac{2\sigma_b}{k_b}\bar{L}^b_i$ acts
only on leg $b$. Identities for the other three-point processes follow
from~\eqref{eq:S3WI} with the crossing rules~\eqref{eq:crossing} applied throughout, and
with the operator on an outgoing leg read from the second column of~\eqref{eq:Qacomms}.

The identity~\eqref{eq:S3WI} takes a more compact form in spinor-helicity variables. It is
convenient to contract the free index with a complex null vector,
\begin{equation}\label{eq:bnull}
b_ib_i = 0, \qquad
b^{a\dot{a}} = \tfrac12 b_i\left(\sigma^i\right)^{a\dot{a}} = \mu^a\bar{\mu}^{\dot{a}},
\end{equation}
which parametrizes the boosts by a pair of spinors $\mu,\bar\mu$. It is a nontrivial computation (see Appendix~\ref{app:boostspinor}) to show that the generator on a single
leg then reduces to a second-order operator on the spinor variables,
\begin{equation}\label{eq:Kspinor}
b_i\left(\bar{\mathcal{K}}_i - \frac{2\sigma}{k}\bar{L}_i\right)
= 2\,k^{-1/2}\,\mu^a\bar{\mu}^{\dot{a}}\,
  \frac{\partial^2}{\partial\lambda^a\partial\bar{\lambda}^{\dot{a}}}\;k^{1/2},
\end{equation}
the conjugation by $k^{1/2}$ accounting for the normalization of the external states
in~\eqref{eq:modeexpansion}. This is the momentum-space special conformal generator
of~\cite{Maldacena:2011nz} up to that factor. Writing
\begin{equation}\label{eq:Dspinor}
\mathcal{D}_{\bf b} \equiv 2\sum_{j=1}^{3}\mu^a\bar{\mu}^{\dot{a}}\,
  \frac{\partial^2}{\partial\lambda_j^a\partial\bar{\lambda}_j^{\dot{a}}} ,
\end{equation}
and contracting~\eqref{eq:S3WI} with ${\bf b}$, the identity becomes
\begin{equation}\label{eq:S3WIspinor}
\left(k_1k_2k_3\right)^{-1/2}\,\mathcal{D}_{\bf b}
\left[\left(k_1k_2k_3\right)^{1/2}\tilde{S}^{+++}_{3}\right]
= \frac{i\,g\,f^{a_1a_2a_3}}{\left(k_1k_2k_3\right)^{3/2}}
\left[\langle\mu\bar{2}\rangle\langle\bar{\mu}\bar{2}\rangle
      \langle\bar{1}\bar{3}\rangle^{2}\,\frac{k_1-k_3}{k_2^{2}}
      + {\rm cyc.}\right],
\end{equation}
with $\tilde{S}^{+++}_{3}$ as in~\eqref{eq:S3helicity} and the remaining helicity
configurations obtained by the crossing rules described in Appendix~\ref{app:spinorhelicity}. This identity was first obtained by Maldacena and Pimentel~\cite{Maldacena:2011nz}, who act with
the special conformal generator on late-time correlators of conserved currents; there the
inhomogeneous term comes from the longitudinal part of the current, fixed by the Ward identity for
current conservation. Here it arises instead from the compensating gauge transformation~\eqref{eq:compensating}, as a boundary term at $\mathcal{I}^+$. The
inhomogeneous term is thus not an artifact of the S-matrix or of Coulomb gauge; it is already
present, with the same form, in the boundary Ward identity for conserved currents.

The four-point amplitude obeys an inhomogeneous identity of the same kind. The compensator
enters through the $\mathcal{O}(g^2)$ piece of $\left[Q_{K_i},U\right]$, which now receives
contributions from $\lambda^{a(1)}_i$ as well as from the quartic and Coulomb vertices, and
the resulting boundary term at $\mathcal{I}^+$ does not seem to organize itself into anything as
compact as~\eqref{eq:S3WI}. We have verified that it is nonzero for all helicites at general kinematics.

\section{Flat limit}\label{sec:flatlimit}

A useful check when computing de Sitter amplitudes is that they should reduce to their Minkowski counterparts in some regime insensitive to the curvature of de Sitter. The problem we discuss in this paper is pure Yang-Mills in the expanding Poincar\'e patch -- a conformally invariant theory in a conformally flat spacetime. All factors of the de Sitter radius $\ell$ cancel in the action and amplitudes, so there is no parameter to take a limit of. Additionally, we cannot zoom in on a region small compared with the Hubble radius by scaling the momenta, since that operation is a dilation and the dilation is an exact symmetry of the patch. The only structures that differentiate Yang-Mills in the Poincar\'e patch from the usual Minkowski picture are the symmetries of the spacetime~\eqref{eq:ppMABs} and the range of the time integral~\eqref{eq:etaint}.

It is the latter that lets us extract the flat amplitudes; they show up as the residue of the
total-energy pole. The difference between~\eqref{eq:Sfinal} and the corresponding Minkowski computation is its range: the mode functions are the flat ones by
conformal invariance, and the vertices~\eqref{eq:verticesexplicit} are the flat ones, so the
integrands agree and the patch differs only in ending at $\eta = 0$ rather than continuing to
$\eta = +\infty$. In Minkowski space, the time integral would give
\begin{equation}\label{eq:etaintflat}
\int_{-\infty}^{\infty}\! d\eta\; e^{-ik_T\eta} = 2\pi\,\delta(k_T),
\end{equation}
the statement of energy conservation, whereas~\eqref{eq:etaint} gives $i/(k_T+i\epsilon)$.
Restoring the prescription suppressed after~\eqref{eq:etaint} and using
\begin{equation}\label{eq:sokhotski}
\lim_{\epsilon\to0^+}\frac{1}{k_T+i\epsilon}
 = {\rm p.v.}\!\left(\frac{1}{k_T}\right) - i\pi\,\delta(k_T),
\end{equation}
we see that the Minkowski delta function is still present, accompanied by a principal value with no flat-space counterpart, and this is the imprint of the boundary at $\mathcal{I}^+$. The flat space amplitude is obtained as the residue of the $1/k_T$ pole. The origin of the pole at $k_T = 0$ is the lower endpoint of the time integral; the integrand is finite at $\mathcal{I}^+$, so nothing diverges there. The residue is therefore built entirely from the far past of the patch, where the external legs have wavelengths small compared with the de Sitter radius: a mode of co-moving momentum $k_b$ has physical wavelength $\lambda_b = 2\pi\ell/|k_b\eta|$, so $\lambda_b/\ell\to0$ as $\eta\to-\infty$
and every external leg is agnostic to the expansion. What the residue extracts is the amplitude for a process taking place at wavelengths small compared with $\ell$, which
is the Minkowski amplitude. 

The residues of the three-point amplitudes~\eqref{eq:S3helicity} follow immediately, since
all of the $k_T$ dependence sits in the scalar prefactor: 
\begin{equation}\label{eq:S3residues}
\begin{aligned}
\lim_{k_T \to 0}k_T\,\tilde{S}_{3}^{++-} & = \frac{i\,g\,f^{a_1a_2a_3}}{\sqrt{k_1k_2k_3}}\,
   \frac{\langle\bar{1}\bar{2}\rangle^3}{\langle\bar{1}\bar{3}\rangle\langle\bar{2}\bar{3}\rangle},
\\[4pt]
\lim_{k_T\to0}k_T\,\tilde{S}_{3}^{+--}  & = \frac{i\,g\,f^{a_1a_2a_3}}{\sqrt{k_1k_2k_3}}\,
   \frac{\langle 23\rangle^3}{\langle 21\rangle\langle 31\rangle},
\end{aligned}
\end{equation}
while $\tilde{S}^{+++}_{3}$ and $\tilde{S}^{---}_{3}$ are regular at $k_T = 0$ and
have no residue at all. These are the Minkowski three-point amplitudes: the two mixed
configurations are the anti-MHV and MHV structures, and the all-plus and all-minus
configurations vanish on shell in flat space. Up to the external normalization
$\prod_b(2k_b)^{-1/2}$ carried by our asymptotic states, \eqref{eq:S3residues} is Parke-Taylor at three points. Note that away from $k_T = 0$, the all-plus and all-minus amplitudes are nonzero. Their vanishing in Minkowski space is a consequence of four-momentum conservation.
It is worth asking what this means operationally. The ratio $k_T/k_b$ sets the longest wavelength of leg $b$, in units of $\ell$, in the region that generates the pole. The singular
part of~\eqref{eq:etaint} comes from $|\eta| \gtrsim 1/k_T$, and at conformal time $\eta$ the
wavelength of leg $b$ relative to the de Sitter radius is $\lambda_b/\ell \sim 1/|k_b\eta|$, so throughout the region responsible
$\lambda_b/\ell \lesssim k_T/k_b$. For $3\to0$ this
ratio is bounded below. Momentum conservation and the triangle inequality give
$k_T \geq 2\max_b k_b$, so $k_T/k_b \geq 2$ for every leg and no choice of real external momenta
keeps the wavelengths small compared with $\ell$ throughout the region responsible. The pole is reached only on complexified kinematics,
and $3\to0$ is intrinsically sensitive to wavelengths of order $\ell$, as it must be for something that flat
space forbids. In the crossed channels the situation is different. For $2\to1$,
$k_T = k_1 + k_2 - k_3$ vanishes in the collinear configuration, so tuning the two incoming
momenta towards collinearity drives $k_T/k_b \to 0$: every leg then has $\lambda_b \ll \ell$ throughout the
region responsible, which is the flat-space regime. The collinear limit is precisely where flat massless three-point kinematics
degenerates, and in this sense~\eqref{eq:S3residues} represent the near-collinear behavior of the
$2\to1$ amplitude.

At four points the same statement holds for the partial
amplitudes~\eqref{eq:A1234}. Since the whole $k_T$ dependence of $\tilde{S}_{4}$ sits in
the prefactor of~\eqref{eq:S4color}, the pole is simple and taking the residue amounts to
evaluating $A$ at $k_T = 0$. For $4\to0$, as for $3\to0$, this locus lies on complexified kinematics, so the residue is a property of $\tilde{S}_{4}$ as an analytic function, fixed by its values in the physical region, rather than of any physical configuration. Setting $k_T = 0$ fuses the two partial-energy denominators in each block into a single flat propagator,
\begin{equation}\label{eq:fuse}
\frac{1}{2P_s}\left(\frac{1}{k_1+k_2-P_s} + \frac{1}{k_3+k_4-P_s}\right)
\;\xrightarrow{\;k_T\to0\;}\;
\frac{1}{\left(k_1+k_2\right)^2 - P_s^2}.
\end{equation}
What remains is
\begin{equation}\label{eq:A1234residue}
\begin{aligned}
A(1^-,2^-,3^+,4^+)\big|_{k_T=0} &= 2\,
  \frac{\langle 12\rangle^4}
       {\langle 12\rangle\langle 23\rangle\langle 34\rangle\langle 41\rangle},
\end{aligned}
\end{equation}
which is the flat-space color-ordered MHV four-gluon amplitude in the all-incoming convention. The remaining configurations, those with
fewer than two legs of either helicity, vanish at $k_T = 0$.

We may also look at the $k_T = 0$ behavior of the Ward identity. The right hand side
of~\eqref{eq:S3WI} is regular at $k_T = 0$, so the left hand side must also be regular. The
amplitude $\tilde{S}_{3}$, however, has a simple pole for the MHV and anti-MHV helicity
configurations, and since the boost operator is second order in derivatives, the left hand
side could generate poles up to order $k_T^{-3}$. To see how these poles cancel, we write
$(k_1k_2k_3)^{1/2}\tilde{S}_3 = N/k_T$, with $N$ regular at $k_T = 0$. Contracted with
${\bf b}$, the left hand side of~\eqref{eq:S3WI} is then $(k_1k_2k_3)^{-1/2}D_{\bf b}(N/k_T)$, and
\begin{equation}\label{eq:DbNkT}
D_{\bf b}\,\frac{N}{k_T}=\frac{D_{\bf b}N}{k_T}+\frac{2\,\mathcal{J}_{\bf b}N}{k_T^2},
\qquad
\mathcal{J}_{\bf b}=\sum_{n=1}^{3}\Big(\langle\lambda_n,\mu\rangle\,\mu^a\frac{\partial}{\partial\lambda_n^a}
+\langle\mu,\bar\lambda_n\rangle\,\mu^{\dot a}\frac{\partial}{\partial\bar\lambda_n^{\dot a}}\Big).
\end{equation}
Here $D_{\bf b}$ acts on $N$ as a function of unconstrained spinors; we write the terms that remain after imposing three-momentum conservation. The operator $\mathcal{J}_{\bf b}$ is the
flat space Lorentz boost along ${\bf b}$. It appears because $D_{\bf b}$ coincides with the flat
space special conformal generator along ${\bf b}$~\cite{Witten:2003nn}, which does not commute
with the flat space energy: $[D_{\bf b},k_T]=-2\mathcal{J}_{\bf b}$.

Regularity of the left hand side at $k_T = 0$ therefore amounts to two conditions,
\begin{equation}
\mathcal{J}_{\bf b}N\big|_{k_T=0}=0,\qquad
\big(D_{\bf b}N+2\,\partial_{k_T}\mathcal{J}_{\bf b}N\big)\big|_{k_T=0}=0,
\end{equation}
which remove the double and the simple pole respectively. Both depend on $N$ only through its
value at $k_T = 0$, since shifting $N\to N+k_TG$ shifts $N/k_T$ by the regular function $G$.
At $k_T = 0$, $N$ is proportional to the flat space three-gluon amplitude $A_3$, which, written
in terms of $\langle\lambda_m,\lambda_n\rangle$ or $\langle\bar\lambda_m,\bar\lambda_n\rangle$
alone, is annihilated by both $D_{\bf b}$ and $\mathcal{J}_{\bf b}$, so both conditions hold.
Equivalently, the two conditions state that the coefficients of $\delta'(k_T)$ and
$\delta(k_T)$ in $D_{\bf b}(\delta(k_T)N)$ vanish, so the regularity of the left hand side
of~\eqref{eq:S3WI} relies on the Lorentz and special conformal invariance of $A_3$. For the de
Sitter amplitude itself, $D_{\bf b}N$ does not vanish at $k_T=0$, and the simple pole cancels
between the two terms of~\eqref{eq:DbNkT}. What remains at $k_T = 0$ is the finite part
of~\eqref{eq:DbNkT}. It depends on the amplitude beyond its residue, and it matches the right
hand side of~\eqref{eq:S3WI}, which is non-zero there. The poles are thus fixed entirely by flat
space physics, while the de Sitter content of the Ward identity resides in its regular part.

\section{Summary and conclusions}\label{sec:conclusions}

In Minkowski spacetime, quantum field theory relies on Poincar\'e invariance. All inertial observers and their time-like geodesics are related by symmetry transformations. Their history is governed by canonical Hamiltonians. In the expanding universe, described as the Poincar\'e patch of de Sitter spacetime, the metrics are invariant under $SO(1,4)$ transformations.
Time-like geodesics fall into two classes: they may have  finite or infinite past, as measured in the units of observer's proper time. Only the observers with infinite past can access the asymptotic ``in'' states of quantum field theory, therefore the S-matrix formalism developed in this work has been focused on the congruence of ``eternal'' geodesics. This congruence is not invariant under the full $SO(1,4)$ because the three boost generators (which become the special conformal generators at the future boundary $\mathcal{I}^+$) transform eternal geodesics into ``newborns'' with finite past. The remaining symmetry is the 
similitude subgroup ${\rm Sim}(3)$ which is composed of three translations, three rotations, and the dilation (uniform scaling). Proper time translations along eternal geodesics are generated by dilations. To construct the scattering amplitudes, we followed the strategy of~\cite{Taylor:2024vdc} and employed the dilation generator as the time evolution operator. In this way, we obtained the curved space generalization of Dyson's formula in the expanding Poincar\'e patch for computing scattering amplitudes order by order in perturbation theory. In this work, we applied it to three- and four-gluon amplitudes at the tree level.

Since de Sitter metrics are conformally flat and classical Yang-Mills theory is conformally invariant, the amplitudes do not depend on de Sitter radius at tree level. Three-gluon amplitudes have a particularly simple form and can be expressed in spinor helicity notation. Unlike in Minkowski spacetime, they are nonvanishing,  because the ${\rm Sim}(3)$ group does not impose as strong kinematic constraints as the Poincar\'e invariance. We analyzed the Ward identities associated to the three broken boost (special conformal) generators and traced the origin of the offending terms to the variations of Lagrangian density. Although the physical meaning of scattering amplitudes is different from the cosmological correlators, the three-gluon expressions are given by the same functions of the kinematic variables. Four-gluon amplitudes are given by rather complicated expressions.

We showed that similarly to cosmological correlators, the scattering amplitudes are singular when the total energy of the particles before the collision is equal to the final energy, as expected by the energy conservation law in the flat limit. Flat amplitudes are recovered as the residues of the respective pole singularities. We found that in  this limit, the interactions are bound to the subhorizon, approximately flat region of de Sitter spacetime. In general, however, horizon scale processes can contribute to the amplitudes. For that reason, we do not expect simple expressions for multi-gluon amplitudes expressed in the momentum representation  on the Poincar\'e patch, even by using  the spinor helicity techniques. It is possible though, that they can be simplified by changing the basis of the Hilbert space, or by connecting them to the amplitudes studied in global de Sitter spacetime \cite{Taylor:2024vdc,Taylor:2025spp,Taylor:2026ase}.

\section*{Acknowledgments}
AL thanks Sophie (cat) for her continued moral support and willingness to be included in thought experiments. This work was supported in part by NSF grants PHY-2209903 and  PHY-2609835, the Simons Collaboration on Celestial Holography, and by the MAESTRO grant no. 2024/54/A/ST2/00009
funded by the National Science Centre, Poland. 
Any opinions, findings, and conclusions or
recommendations expressed in this material are those of the authors and do not necessarily
reflect the views of the National Science Foundation.

\appendix

\section{Spinor-helicity variables}\label{app:spinorhelicity}

Here we collect definitions and conventions for the spinor-helicity variables. Just as in flat space, each three-momentum is promoted to a null four-vector and factorized into spinors. The difference is that the full
${\rm SL}(2,\mathbb{C})\times{\rm SL}(2,\mathbb{C})$ is broken to the diagonal $SU(2)$, the
double cover of the rotation group $SO(3)$. Dotted and undotted spinors therefore transform
in the same way, and may be contracted with one another using $\epsilon_{\dot{a}b}$. The two types of spinor
remain distinct, but the barred/unbarred
distinction no longer obstructs contraction.

\subsection*{Conventions}\label{app:SHconventions}

The $\sigma$ matrices are
\begin{equation}
(\sigma^{\mu})^a{}_b = \left( -\delta^{a}{}_b,\, (\sigma^i)^a{}_b\right),
\end{equation}
with $\sigma^i$ the Pauli matrices, and indices are raised and lowered by the antisymmetric
\begin{equation}
\epsilon_{ab} = \begin{pmatrix}
0 & 1 \\ -1 & 0
\end{pmatrix}, \qquad \epsilon^{ab} = \begin{pmatrix}
0 & -1 \\ 1 & 0
\end{pmatrix}
\end{equation}
as
\begin{equation}
(\sigma^{\mu})_{ab} \equiv -\epsilon_{ac}(\sigma^{\mu})^c{}_b, \qquad
(\sigma^{\mu})^{ab} \equiv -(\sigma^{\mu})^a{}_c\,\epsilon^{cb}.
\end{equation}
Dotted indices are decorative,
\begin{equation}
\epsilon_{ab} = \epsilon_{\dot{a}b} = \epsilon_{a\dot{b}} = \epsilon_{\dot{a}\dot{b}},
\end{equation}
and for any two spinors $X, Y$ we define the angle bracket
\begin{equation}\label{eq:bracket}
\langle XY\rangle \equiv X^a\epsilon_{ab}Y^b = -\langle YX \rangle.
\end{equation}

For a state of momentum ${\bf k}$ we define the null four-vector
\begin{equation}\label{eq:ktilde}
\tilde{k}^{\mu} = \left(\tilde{k}, \tilde{{\bf k}}\right) = \pm\left(k, {\bf k}\right),
\qquad k = |{\bf k}|,
\end{equation}
with $+$ for an incoming state and $-$ for an outgoing one. Its bispinor factorizes as in
flat space,
\begin{equation}\label{eq:spinorfactorize}
k^{a\dot{b}} = \frac12\left(\tilde{k}\,(\sigma^0)^{a\dot{b}}
+ \tilde{k}_i\,(\sigma^i)^{a\dot{b}}\right) = \lambda^a\bar{\lambda}^{\dot{b}},
\qquad
\langle \lambda \bar{\lambda} \rangle = -\tilde{k}, \qquad
\langle \lambda \lambda \rangle = \langle \bar{\lambda}\bar{\lambda}\rangle = 0,
\end{equation}
with explicit components
\begin{equation}\label{eq:spinorcomponents}
\lambda = \left(\frac{-\tilde{k}_1+i\tilde{k}_2}{\sqrt{2(\tilde{k}+\tilde{k}_3)}},\
\sqrt{\frac{\tilde{k}+\tilde{k}_3}{2}}\right)^T, \qquad
\bar{\lambda} = \left(\sqrt{\frac{\tilde{k}+\tilde{k}_3}{2}},\
\frac{\tilde{k}_1+i\tilde{k}_2}{\sqrt{2(\tilde{k}+\tilde{k}_3)}}\right).
\end{equation}
For an incoming state these obey the reality condition
\begin{equation}\label{eq:reality}
\left(\bar{\lambda}^{\dot{a}}\right)^* = \epsilon_{\dot{a}b}\,\lambda^b.
\end{equation}
With
several legs we label spinors by their momenta, writing
$\langle kp\rangle \equiv \langle\lambda_k\lambda_p\rangle$,
$\langle\bar{k}p\rangle \equiv \langle\bar{\lambda}_k\lambda_p\rangle$, and so on.

\subsection*{Polarization vectors}

In flat space the polarization vectors are
\begin{equation}\label{eq:xiflat}
\xi^{-\,a\dot{b}} = \frac{\lambda^a\bar{r}^{\dot{b}}}{\langle\bar{r}\bar{\lambda}\rangle},
\qquad
\xi^{+\,\dot{a}b} = \frac{\bar{\lambda}^{\dot{a}}r^{b}}{\langle r\lambda\rangle},
\end{equation}
where $\bar{r},r$ are reference spinors, required only to be non-proportional to
$\bar{\lambda},\lambda$ respectively. Changing them shifts $\xi^{\pm}$ by a multiple of
$k^{a\dot{b}}$, which is a gauge transformation and drops out of physical amplitudes. In
flat space $\bar{r}$ must be barred and $r$ unbarred; here the diagonal embedding removes
that restriction, and we may take $\bar{r}\to\lambda$ and $r\to\bar{\lambda}$, giving

\begin{equation}\label{eq:xidS}
\xi^{-\,ab} = \frac{\lambda^a\lambda^b}{\langle\lambda\bar{\lambda}\rangle},
\qquad
\xi^{+\,\dot{a}\dot{b}} = \frac{\bar{\lambda}^{\dot{a}}\bar{\lambda}^{\dot{b}}}{\langle\bar{\lambda}\lambda\rangle},
\qquad \langle\bar{\lambda}\lambda\rangle = \tilde{k}.
\end{equation}
Reconstructing the polarization vectors as
$\xi^{\pm}_{\mu} = (\sigma_{\mu})_{ab}\,\xi^{\pm\,ba}$, the temporal component vanishes,
$\xi^{\pm}_0 = 0$, and the spatial components are related to the polarization vectors
$\epsilon_{\sigma}({\bf k})$ of~\eqref{eq:modeexpansion} by
\begin{equation}\label{eq:xitoeps}
\xi^{\pm}_i = \sqrt{2}\,\epsilon_{\pm,i}({\bf k}).
\end{equation}
The factorization~\eqref{eq:spinorfactorize} fixes $\lambda,\bar{\lambda}$ only up to
$\lambda\to w\lambda$, $\bar{\lambda}\to w^{-1}\bar{\lambda}$, with $|w| = 1$
by~\eqref{eq:reality}, under which $\xi^{\pm}\to w^{\mp 2}\xi^{\pm}$.

Products of polarization vectors and momenta are
\begin{equation}\label{eq:xiproducts}
\xi^+_k \cdot \xi^+_p = -2\frac{\langle \bar{k}\bar{p}\rangle^2}{\tilde{k}\tilde{p}},
\qquad
\xi^-_k \cdot \xi^-_p = -2\frac{\langle kp\rangle^2}{\tilde{k}\tilde{p}},
\qquad
\xi^+_k \cdot \xi^-_p = 2\frac{\langle \bar{k}p\rangle^2}{\tilde{k}\tilde{p}},
\end{equation}
\begin{equation}
\tilde{{\bf k}} \cdot \xi^+_p = -\frac{2\langle k\bar{p}\rangle\langle\bar{k}\bar{p}\rangle}{\tilde{p}},
\qquad
\tilde{{\bf k}} \cdot \xi^-_p = \frac{2\langle kp\rangle\langle\bar{k}p\rangle}{\tilde{p}}.
\end{equation}
For three legs satisfying momentum conservation,
\begin{equation}
\tilde{{\bf k}} + \tilde{{\bf p}} + \tilde{{\bf q}} = 0,
\end{equation}
with total energy $k_T = \tilde{k} + \tilde{p} + \tilde{q}$, we also have
\begin{equation}\label{eq:threepointidentities}
\langle k\bar{p} \rangle = \frac{2}{k_T}\langle kq \rangle \langle \bar{q}\bar{p} \rangle,
\qquad
\langle kp\rangle \langle \bar{k}\bar{p} \rangle = \frac14 k_T(k_T - 2\tilde{q}).
\end{equation}

All of these hold for any assignment of incoming and outgoing legs.

\subsection*{Crossing rules}

The crossing rules~\eqref{eq:crossing} act simply on the spinor variables. Since
$\tilde{k}^{\mu} \to -\tilde{k}^{\mu}$, the bispinor~\eqref{eq:spinorfactorize} flips sign,
and the branch fixed by the Bunch--Davies prescription distributes that sign symmetrically
between the two spinors:
\begin{equation}\label{eq:spinorcrossing}
\lambda_b \to i\,\lambda_b, \qquad \bar{\lambda}_b \to i\,\bar{\lambda}_b,
\qquad \sigma_b \to -\sigma_b,
\end{equation}
consistent with $\langle\lambda_b\bar{\lambda}_b\rangle = -\tilde{k}_b$.
For an outgoing state the reality condition~\eqref{eq:reality} therefore acquires a sign,
$(\bar{\lambda}^{\dot{a}})^* = -\epsilon_{\dot{a}b}\,\lambda^b$.
The polarization bispinors~\eqref{eq:xidS} are unchanged, since the factors of $i$ cancel
between numerator and denominator, so the helicity flip in~\eqref{eq:crossing} only
relabels which physical helicity the outgoing particle carries. In practice, crossing leg
$b$ in an all-incoming amplitude multiplies each bracket containing $b$ by $i$ and each
explicit factor of $\tilde{k}_b$ by $-1$.
\section{Spinor form of the boost generator}\label{app:boostspinor}

We outline the derivation of~\eqref{eq:Kspinor} for an incoming leg, with
$\langle\lambda\bar\lambda\rangle=-k$. The operator
$\bar{\mathcal{K}}_i-\frac{2\sigma}{k}\bar L_i$ of~\eqref{eq:Qacomms} acts on functions of
$(\lambda,\bar\lambda)$ of little-group weight $w=-2\sigma$, which we express through
\begin{equation}
\mathcal{E}\equiv\lambda\cdot\partial_\lambda,\qquad
\bar{\mathcal{E}}\equiv\bar\lambda\cdot\partial_{\bar\lambda},\qquad
R\equiv\bar\lambda\cdot\partial_\lambda,\qquad
\bar R\equiv\lambda\cdot\partial_{\bar\lambda},\qquad
\mathcal{E}-\bar{\mathcal{E}}=w .
\end{equation}

Expanding $\partial_j\lambda$ and $\partial_j\bar\lambda$ in the
basis $\{\lambda,\bar\lambda\}$ turns the chain rule into
$\partial_j=\gamma_j\mathcal{E}+\bar\gamma_j\bar{\mathcal{E}}+\beta_jR+\bar\beta_j\bar R$, with
$\gamma_j+\bar\gamma_j=k_j/k^2$, $\beta_j=\xi^-_j/2k$ and $\bar\beta_j=-\xi^+_j/2k$. The
connection in~\eqref{eq:Dcov} is $\alpha_{+,j}=\bar\gamma_j-\gamma_j$, and on functions of weight
$w=-2\sigma$ the individual $\gamma_j$, $\bar\gamma_j$ drop out:
\begin{equation}\label{eq:Dbarspinor}
\bar D_j=\frac{k_j}{k^2}\left(\mathcal{E}+\sigma\right)+\beta_jR+\bar\beta_j\bar R .
\end{equation}

Substituting~\eqref{eq:Dbarspinor} into~\eqref{eq:boostpiece},
and reducing with the commutators of $\mathcal{E},\bar{\mathcal{E}},R,\bar R$, the weight
condition, and their action on $k_i$, $\beta_i$, $\bar\beta_i$ (e.g.\ $Rk_i=-2k^2\bar\beta_i$,
$R\beta_i=-k_i/k^2$), together with $\bar L_i=-k(\beta_iR-\bar\beta_i\bar R)$, gives
\begin{equation}\label{eq:Kbarspinor}
\bar{\mathcal{K}}_i-\frac{2\sigma}{k}\bar L_i
=\frac{k_i}{k^2}\left(\mathcal{E}\bar{\mathcal{E}}+R\bar R\right)
+2\beta_i\bar{\mathcal{E}}R+2\bar\beta_i\mathcal{E}\bar R
-\beta_iR-\bar\beta_i\bar R+(2w-1)\frac{k_i}{4k^2}.
\end{equation}
The terms linear in $\mathcal{E}+\bar{\mathcal{E}}$ cancel, and because $w=-2\sigma$ so does the
helicity dependence of the $R$ and $\bar R$ terms.

Writing $\mu=c_\mu\lambda+d_\mu\bar\lambda$ and
$\bar\mu=c_{\bar\mu}\lambda+d_{\bar\mu}\bar\lambda$, with $c_\nu=-\langle\nu\bar\lambda\rangle/k$
and $d_\nu=\langle\nu\lambda\rangle/k$, gives $\mu\cdot\partial_\lambda=c_\mu\mathcal{E}+d_\mu R$
and $\bar\mu\cdot\partial_{\bar\lambda}=d_{\bar\mu}\bar{\mathcal{E}}+c_{\bar\mu}\bar R$. Since
$\langle\mu\bar\mu\rangle=0$, the coefficients are fixed by~\eqref{eq:bnull},
\begin{equation}
c_\mu d_{\bar\mu}=d_\mu c_{\bar\mu}=\frac{{\bf b}\cdot{\bf k}}{2k^2},\qquad
c_\mu c_{\bar\mu}={\bf b}\cdot\bar\beta,\qquad
d_\mu d_{\bar\mu}={\bf b}\cdot\beta ,
\end{equation}
and commuting the operators through $k^{1/2}$ and the coefficients gives
\begin{equation}
\begin{aligned}
2k^{-1/2}\mu^a\bar\mu^{\dot a}\frac{\partial^2}{\partial\lambda^a\partial\bar\lambda^{\dot a}}\,k^{1/2}
&=\frac{{\bf b}\cdot{\bf k}}{k^2}\left(\mathcal{E}\bar{\mathcal{E}}+R\bar R\right)
+2({\bf b}\cdot\beta)\,\bar{\mathcal{E}}R+2({\bf b}\cdot\bar\beta)\,\mathcal{E}\bar R\\
&\quad-({\bf b}\cdot\beta)R-({\bf b}\cdot\bar\beta)\bar R+(2w-1)\frac{{\bf b}\cdot{\bf k}}{4k^2},
\end{aligned}
\end{equation}
which is ${\bf b}$ contracted with~\eqref{eq:Kbarspinor}. Since $w$ was kept general,
\eqref{eq:Kspinor} holds for both helicities. The steps are identical for showing the relation holds for an outgoing leg.

\end{document}